\documentclass[twocolumn,showpacs,aps,prb]{revtex4}
\usepackage{hyperref}
\usepackage{amsmath,amssymb}
\usepackage{dcolumn}
\usepackage{epsfig}

\begin{document}

\title{Influence of Point-like Disorder on the Guiding of Vortices
and the Hall Effect in a Washboard Planar Pinning Potential}

\author{Valerij~A.~Shklovskij}
\address
{Institute of Theoretical Physics, National Science Center-Kharkov
Institute of Physics and Technology, 61108, Kharkov,
Ukraine;\\Kharkov National University, Physical Department, 61077,
Kharkov, Ukraine}
\author{Oleksandr~V.~Dobrovolskiy}
\address {Kharkov National University, Physical Department, 61077, Kharkov,
Ukraine}\date{\today}
\begin{abstract}
Explicit current-dependent expressions for anisotropic
longitudinal and transverse nonlinear magnetoresistivities are
represented and analyzed on the basis of a Fokker-Planck approach
for two-dimensional single-vortex dynamics in a washboard pinning
potential in the presence of point-like disorder.  Graphical
analysis of the resistive responses is presented both in the
current-angle coordinates and in the rotating current scheme. The
model describes nonlinear anisotropy effects caused by the
competition of point-like (isotropic) and anisotropic pinning.
Nonlinear guiding effects are discussed and the critical current
anisotropy is analyzed.  Gradually increasing the magnitude of
isotropic pinning force this theory predicts a gradual decrease of
the anisotropy of the magnetoresistivities. The physics of
transition from the new scaling relations for anisotropic Hall
resistance in the absence of point-like pins to the well-known
scaling relations for the point-like disorder is elucidated. This
is discussed in terms of a gradual isotropizaton of the guided
vortex motion, which is responsible for the existence in a
washboard pinning potential of new (with respect to magnetic field
reversal) Hall voltages. It is shown that whereas the Hall
conductivity is not changed by pinning, the Hall resistivity can
change its sign in some current-angle range due to presence of the
competition between \emph{i}- and \emph{a}-pins.
\end{abstract}
 \maketitle

\section{INTRODUCTION}
The importance of flux-line pinning in preserving the
superconductivity in a magnetic field has been generally
recognized since the discovery of type-II superconductivity. But
till now the mechanism of flux-line pinning and creep in
superconductors (and particularly in the high-\emph{$T_c$}
superconductors (HTSC's)) is still a matter of controversy and
great current interest, especially in the cases of strong
competition between different types of pins.

One of the open issues in the field is the influence of
\emph{isotropic} point-like disorder on the vortex dynamics in the
\emph{anisotropic} washboard planar pinning potential (PPP) for
the case of arbitrary orientation of the transport current with
respect to the PPP "channels" where the \emph{guiding of vortices
}can be realized. The importance of this issue may be
substantiated  by ubiquitous presence of point-like pins in those
high- and low-\emph{$T_c$} superconductors which were used so far
for resistive measurements of the guided vortex motion$^{1-9}$.

The first attempt to discuss the influence of isotropic point-like
disorder on the guiding of vortices was made by Niessen and
Weijsenfeld$^1$ still in 1969. They studied guided motion \emph{in
the flux flow regime} by measuring transverse voltages of
cold-rolled sheets of a Nb-Ta alloy for different magnetic fields
\emph{H}, transport current densities \emph{J}, temperatures
\emph{T}, and different angles $\alpha$ between the rolling and
current direction. The (\emph{H,J,T},$\alpha$)-dependences of the
cotangent of the angle $\beta$ between the average vortex velocity
$\langle \textbf{v}\rangle$ and the vector \textbf{J} direction
were presented. For the discussion, a simple theoretical model was
suggested, based on the assumption that vortex pinning and guiding
can be described in terms of an isotropic pinning force ${\bf
F}_p^i$ plus a pinning force ${\bf F}_p^a$ with a fixed direction
which was perpendicular to the rolling direction. The
experimentally observed dependence of the transverse and
longitudinal voltages on the magnetic field \emph{in the flux flow
regime} as a function of the angle $\alpha$ was in agreement with
this model.

Unfortunately, in spite of the correct description of a geometry
of the motive forces of a problem (see below Fig.~1) it was
impossible within the flux flow approach$^1$ to calculate
theoretically the \emph{nonlinear }(\emph{J,~T},
$\alpha$)-dependences of the average pinning forces $\langle{\bf
F}_p^i\rangle$ and $\langle{\bf F}_p^a\rangle$ which determine the
experimentally observed cot$\beta(J,T,\alpha)$ dependences.

The \emph{nonlinear guiding }problem was exactly solved at first
only for the washboard PPP (i.e. for ${\bf F}_p^i=0 $) within the
framework of the two-dimensional  single-vortex stochastic model
of anisotropic pinning based on the Fokker-Planck equation with a
concrete form of the pinning potential$^{10,11}$. Two main reasons
stimulated these theoretical studies. First, in some HTCS's twins
can easily be formed during  the crystal growth$^{2-5,8}$. Second,
in layered HTCS's the system of interlayers between parallel
\emph{ab}-planes can be considered as a set of unidirectional
planar defects which provoke the intrinsic pinning of
vortices$^{12}$.

Rather simple formulas were derived$^{11}$ for the experimentally
observable \emph{nonlinear} even$(+)$ and odd$(-)$ (with respect
to the magnetic field reversal) longitudinal and transverse
magnetoresistivities
$\rho_{\|,\perp}^\pm(j,\theta,\alpha,\varepsilon)$ as functions of
the dimensionless transport current density $j,$ dimensionless
temperature $\theta,$ and relative volume fraction
$0<\varepsilon<1$ occupied by the parallel twin planes directed at
an angle $\alpha$ with respect to the current direction. The
$\rho_{\|,\perp}^\pm$-formulas were presented as linear
combinations of the even and odd parts of the function
$\nu(j,\theta,\alpha,\varepsilon)$ which can be considered as the
probability of overcoming the potential barrier of the
twins$^{11}$; this made it possible to give a simple physical
treatment of the nonlinear regimes of vortex motion (see below
item II.C).

Besides the appearance of a relatively large even transverse
$\rho_\perp^+$ resistivity, generated by the guiding of vortices
along the channels of the washboard PPP, explicit expressions for
\emph{two new nonlinear anisotropic Hall resistivities}
$\rho_{||}^-$ \emph{and} $\rho_\perp^-$ were derived and analyzed.
The physical origin of these \emph{odd }contributions caused by
the subtle interplay between even effect of vortex guiding and the
odd Hall effect. Both new resistivities were going to zero in the
linear regimes of the vortex motion (i.e. in the thermoactivated
flux flow (TAFF) and the ohmic flux flow (FF) regimes) and had a
bump-like current or temperature dependence in the vicinity of
highly nonlinear resistive transition from the TAFF to the FF. As
the new odd resistivities arose due to the Hall effect, their
characteristic scale was proportional to the small Hall constant
as for ordinary odd Hall effect investigated earlier$^{10}$. It
was shown$^{11}$ that appearance of these new odd $\rho_{|| , \bot
}^-$ contributions leads to the new specific angle-dependent
"scaling" relations for the PPP which demonstrate the so-called
anomalous Hall behavior in the type-II superconductors.

Here we should to emphasize that the anomalous behavior of the
Hall effect in many high-temperature and in some conventional
superconductors in the mixed state remains one of the challenging
issues in the vortex dynamics$^{5,12,16}$. The problem at issues
includes several remarkable experimental facts: a) the Hall effect
sign reversal in the vortex state with respect to the normal state
at temperatures near $\emph{T}_c$ and for moderate magnetic
fields; b) the Hall resistivity "scaling" relation
$\rho_\perp\sim\rho_\| ^ \beta$ exists with $1\leq\beta\leq2$,
where $\rho_\perp$ is the Hall resistivity and $\rho_\parallel$ is
the longitudinal resistivity; c) the influence of pinning on the
"Hall anomaly" and scaling relation. Assuming that the "bare" Hall
coefficient $\alpha_H$ is constant, two different scaling laws
have been derived earlier theoretically for different pinning
potentials$^{11,17}$. Vinokur et al. have shown$^{17}$ that a
scaling law $\rho_\perp=\delta\rho_\parallel^2$ (where
$\delta=n\alpha_H c^2/B\Phi_0$  is the Hall conductivity,
$n=\pm1$, $c$ is the speed of light, $B$ is the magnetic field and
$\Phi_0$ is the magnetic flux quantum) is the general feature of
any isotropic vortex dynamics with an average pinning force
directed along the average vortex velocity vector. Later it was
shown$^{11}$ that for purely anisotropic \emph{a}-pins that create
a washboard planar pinning potential, the form of corresponding
"scaling" relation is highly anisotropic due to the reason that
pinning force for \emph{a}-pins is directed perpendicular to the
pinning planes. If $\alpha$ is the angle between parallel pinning
planes and direction of the current density vector $\mathbf{j}$,
then for $\alpha=0$ the scaling law has the form
$\rho_\bot=-n(\alpha_H/\eta)\rho_\|$ ($\eta$ is the vortex
viscosity) which was interpreted previously$^{11}$ as a scaling
law with $\beta=1$, whereas for $\alpha=\pi/2$ the scaling
relation is more complex$^{11}$. The $\rho_{\perp}^-$, as it is
shown in this paper, can be presented as a sum of the three
contributions with the different signs. The graphical analysis in
Sec.~III of this paper represents a some range of the
$(\alpha,j)$-values where the theory predicts a nonlinear change
of the $\rho_{\perp}^-$ sign.

Let us consider another specific feature of the purely anisotropic
guiding model$^{10,11}$. From the mathematical viewpoint, the
nonlinear anisotropic problem, as solved in Ref.~11, reduces to
the Fokker-Planck equation of the one-dimensional vortex
dynamics$^{13}$ because the vortex motion is unpinned in the
direction which is parallel to the PPP channels. As a consequence,
a critical current $j_c$ exists only for the direction which is
strictly perpendicular to the PPP channels ($\alpha=0$);
$j_c(\alpha)=0$ for any other direction ($0<\alpha\le \pi/2$).
However, the measurements of the magnetoresistivity show$^{1-8}$
that $j_c(\alpha)>0$ for all $\alpha$ (although $j_c(\alpha)$ may
be anisotropic). So, in spite of some merits of a model with a
washboard PPP, which was the first exactly solvable stochastic
nonlinear model of anisotropic pinning, it cannot describe the
$j_c$-anisotropy of the experimentally measured samples.

Due to this reason later it was suggested$^{14,20}$ another simple
model, which demonstrates this $j_c$-anisotropy for all $\alpha$
on the basis of the bianisotropic pinning potential formed by the
sum of two washboard PPP's in two mutually perpendicular
directions. In contradistinction to the nonlinear model with
uniaxial PPP$^{11}$, this bianisotropic nonlinear model predicts a
$j_c(\alpha)$-anisotropy and relates it to the guiding anisotropy,
describing the appearance of two step-like and two bump-like
singularities in the $\rho_{\|,\perp}^+$ and $\rho_{\|,\perp}^-$
(Hall) resistive responses, respectively. Although several
proposals to realize experimentally this bianisotropic model were
discussed so far$^{14}$, the corresponding experiments, however,
are still absent.

At the same time, the experimental study of vortex dynamics in the
PPP is always accompanied with a presence of a certain level of
point-like disorder. So, as far as the analysis of existing
experimental data is concerned, none of the present theoretical
studies in the limiting cases of purely anisotropic or isotropic
pinning are sufficient. The more general approach is needed.

 The objective of this paper is to present results of a theory
for the calculation of the nonlinear magnetoresistivity tensor at
arbitrary value of competition between point-like and anisotropic
planar disorder for the case of in-plane geometry of experiment.
This approach will give us the experimentally important
theoretical model which demonstrates the $j_c$-anisotropy for all
$\alpha$ and predicts a nonlinear change of the $\rho_\perp^-$
sign at some set of parameters (without change of the Hall
\emph{conductivity}) due a competition of the washboard PPP and a
point-like disorder. \enlargethispage{\baselineskip}

The organization of the article is as follows. In Sec. II we
derive main results of the $i+a$ pinning problem and consider two
main limiting cases of purely $a$- or $i$-pinning. In Sec. III we
represent the graphical analysis of different types of nonlinear
responses, in particular, the $(j,\alpha)$ graphs of the
$\rho_{\parallel,\perp}^\pm$ magnetoresistivities and the
resistive response in a rotating current scheme. In Sec. IV we
conclude with a general discussion of our results.

\section{Main relations}
\subsection{Formulation of the problem.}

 The Langevin equation for
a vortex moving with velocity $\mathbf{v}$ in a magnetic field
$\mathbf{B}=\mathbf{n}B$ ($B\equiv|\mathbf{B}|$,
$\mathbf{n}=n\mathbf{z}$, $\mathbf{z}$~is the unit vector in the
$z$-direction and $n=\pm 1$) has the form
\begin{equation}
\label{F1}
\eta_{0}\mathbf{v}+n\alpha_{H}\mathbf{v}\times\mathbf{z}=\mathbf{F}_{L}+\mathbf{F}_{p}^{a}+\mathbf{F}_{p}^{i}+\mathbf{F}_{th},
\end{equation}
where $\mathbf{F}_{L}=n(\Phi_{0}/c)\mathbf{j}\times\mathbf{z}$ is
the Lorentz force ($\Phi_{0}$ is the magnetic flux quantum, $c$ is
the speed of light), $\mathbf{F}_{p}^{a}=-\nabla U_{p}(x)$ is the
anisotropic pinning force ($U_{p}(x)$ is the washboard planar
pinning potential), $\mathbf{F}_{p}^{i}$ is the isotropic pinning
force, induced by uncorrelated point-like disorder ,
$\mathbf{F}_{th}$ is the thermal fluctuation force, $\eta_{0}$ is
the vortex viscosity, and $\alpha_{H}$ is the Hall constant.

\begin{figure}
\includegraphics{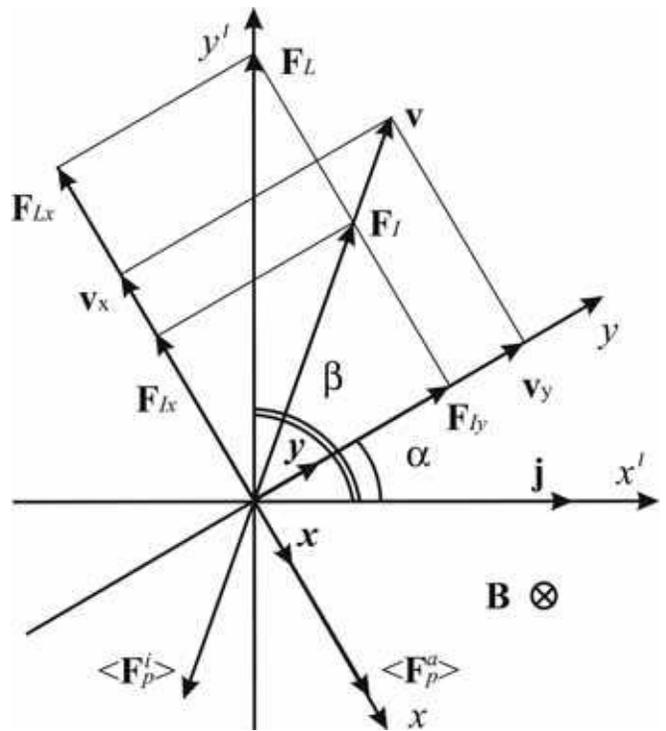}
\caption{System of coordinates $xy$ (with the unit vectors
$\mathbf{x}$ and $\mathbf{y}$) associated with the PPP planes and
the system of coordinates $x'y'$ associated with the direction of
the current density vector $\mathbf{j}$; $\alpha$ is the angle
between the channels of the PPP and $\mathbf{j}$, $\beta$ is the
angle between the average velocity vector of the vortices
$\textbf{v}$ and the vector $\mathbf{j}$; $\mathbf{F}_{L}$ is the
Lorentz force; $<\mathbf{F}_{p}^{i}>$ and $<\mathbf{F}_{p}^{a}>$
are the average isotropic and anisotropic pinning forces,
respectively, $\mathbf{F}_{I}$ is the average effective motive
force for a vortex. Here for simplicity we assume $\epsilon=0$.}
\label{fig1}
\end{figure}

For purely isotropic pinning (i.e. for $\mathbf{F}_{p}^{a}=0$){
Eq.} (\ref{F1}) was earlier solved$^{17}$ for $\mathbf{F}_{th}=0$,
using the fact that
\begin {equation}
\label{F2} \mathbf{F}_{p}^{i}=-\eta_{i}(\upsilon)\mathbf{v},
\end{equation}
where $ \eta_{i}(\upsilon)$ is velocity-dependent viscosity and
$\upsilon\equiv|\mathbf{v}|$.

Below we will show (see Eq.~(8) and item D of Sec. II), that the
solution, obtained in Ref.~17, can be presented in terms of the
probability function of overcoming the effective current- and
temperature-dependent potential barrier of isotropic pinning
$\nu_{i}(F_{I})$, which is simply related to $\eta_{i}(\upsilon)$.

In the absence of point-like disorder (i.e. for
$\mathbf{F}_{p}^{i}=0$)  {Eq.}~(\ref{F1}) was reduced to the
Fokker-Planck equation, which was solved$^{10,11}$, assuming that
the fluctuational force $\mathbf{F}_{th}(t)$ is represented by a
Gaussian white noise, whose stochastic properties are assigned by
the relations
\begin {equation}
\label{F3} \langle F_{th,i}(t)\rangle=0, \, \langle F_{th,i}(t)
F_{th,j}(t') \rangle=2T\eta_{0}\delta_{ij}\delta(t-t'),
\end{equation}
where $T$ is the temperature in energy units.

In what follows we derive the solution of {Eq.} (\ref{F1}), using
for $\mathbf{F}_{p}^{i}$ the assumption (\ref{F2}), which reduces
{Eq.} (\ref{F1}) to the equation
\begin {equation}
\label{F4}
\eta\mathbf{v}+n\alpha_{H}\mathbf{v}\times\mathbf{z}=\mathbf{F}_{L}+\mathbf{F}_{p}^{a}+\mathbf{F}_{th},
\end{equation}
where $\eta=\eta(\upsilon)\equiv\eta_{0}+\eta_{i}(\upsilon)$.
Using the result of Ref.~11, the selfconsistent solution of the
{Eq.} (\ref{F4}) can be represented as
\begin {equation}
\label{F5} \begin{array}{l} \eta(\upsilon)\langle v_{x}
\rangle=F_{a}\nu_{a}(F_{a})/(1+\tilde{\epsilon}^{2}),\\
\\
\eta(\upsilon)\langle v_{y}
\rangle=F_{Ly}+n\tilde{\epsilon}F_{a}\nu_{a}(F_{a})/(1+\tilde{\epsilon}^{2}),
\end{array}
\end{equation}
where $\nu_{a}(F_{a})$ is the probability of overcoming the PPP
under the influence the effective moving force $F_{a}\equiv
F_{Lx}-n\tilde{\epsilon}F_{Ly}$, $F_{Lx}$ and $F_{Ly}$ are the
Lorentz force components acting along the vector $\mathbf{x}$ and
$\mathbf{y}$, respectively, $\tilde{\epsilon}\equiv\epsilon
Z(\upsilon)$, $\epsilon\equiv\alpha_{H}/\eta_{0}$ and
$Z(\upsilon)\equiv\eta_{0}/\eta(\upsilon)$ with an obvious
condition $0\leq Z(\upsilon)\leq1$. {Eqs.} (\ref{F5}) can be
rewritten as
\begin {equation}
\label{F6} \eta(\upsilon)\langle \mathbf{v}
\rangle=\mathbf{F}_{I},
\end{equation}
where  $F_{Ix}$ and $F_{Iy}$ are corresponding right-hand parts of
{Eqs.} (\ref{F5}). From {Eq.} (\ref{F6}) we have
\begin {equation}
\label{F7} \eta(\upsilon)\upsilon=F_{I},
\end{equation}
where $F_{I}\equiv(F_{Ix}^2+F_{Iy}^2)^{1/2}$ and we omitted for
simplicity the symbol of averaging for $\mathbf{v}$. Then from
{Eq}. (\ref{F7}) follows that $\upsilon=\upsilon(F_{I})$ and thus
it is possible to represent $\eta_{i}(\upsilon)$ and $Z(\upsilon)$
in terms of $F_{I}$:
$\eta_{i}(\upsilon)=\eta_{i}[\upsilon=\upsilon(F_{I})]\equiv\tilde{\eta}_{i}(F_{I})$
and
\begin {equation}
\label{F8}
Z(\upsilon)=Z[\upsilon=\upsilon(F_{I})]\equiv\nu_{i}(F_{I}).
\end{equation}
Here $\nu_{i}(F_{I})$ has a physical meaning of the probability to
overcome the effective potential barrier of  isotropic pinning
under the influence of effective ($\upsilon$-dependent through the
$\tilde{\epsilon}$-dependence) force $F_{I}$. Then in terms of the
$\nu_{i}(F_{I})$ {Eq.} (\ref{F6}) takes the selfconsistent form
\begin {equation}
\label{F9} \eta_{0}\mathbf{v}=\nu_{i}(F_{I})\mathbf{F}_{I},
\end{equation}
which can be highly simplified for a small dimensionless Hall
constant\qquad
 $(\epsilon\ll 1)$. Really, in this limit\qquad
 $\tilde{\epsilon}=\epsilon\nu_{i}(F_{i})$, where\qquad
  $F_{i}\equiv  F_{I} (\epsilon=0)$, and the right-hand part of the {Eq.}
(\ref{F6}) becomes $\upsilon$-independent, i.e. is represented
only in terms of the  known quantities. Just in this limit  all
subsequent results of the paper will be discussed.

\subsection{The nonlinear resistivity and  conductivity tensors}

The average electric field induced by the moving vortex system is
given by
\begin {equation}
\label{F10}
\mathbf{E}=(1/c)\mathbf{B}\times\mathbf{v}=n(B/c)(-\upsilon_{y}\mathbf{x}+\upsilon_{x}\mathbf{y}),
\end {equation}
where $\mathbf{x}$ and $\mathbf{y}$ are the unit vectors in $x$-
and $y$-direction, respectively.

From formulas (\ref{F9}) and (\ref{F10}) we find the dimensionless
magnetoresistivity tensor $\hat{\rho}$ (having components measured
in units of the flux-flow resistivity
$\rho_{f}\equiv\Phi_{0}B/\eta_{0}c^{2}$) for the nonlinear law
$\mathbf{E}=\hat{\rho}(j)\mathbf{j}$
\begin {equation}
\label{F11} \begin{array}{l}\hat{\rho}=
\left(\begin {array}{cc} \rho_{xx}& \rho_{xy}\\
\rho_{yx}& \rho_{yy} \end {array}\right )=\\
\qquad{}\\
\qquad{} \left(\begin {array}{cc}
\nu_{i}(F_{I})&-n\epsilon\nu_{i}^2(F_{i})\nu_{a}(F_{Lx})\\
n\epsilon\nu_{i}^2(F_{i})\nu_{a}(F_{Lx})&\nu_{i}(F_{I})\nu_{a}(F_{a})\end
{array}\right ).
\end{array}
\end {equation}

 The conductivity tensor $\hat{\sigma}$ (the components of which
are measured in units of $1/\rho_{f}$), which is the inverse of
the tensor $\hat{\rho}$, has the form
\begin {equation}
\label{F12} \begin{array}{l}\hat{\sigma}=\left(\begin {array}{cc} \sigma_{xx}& \sigma_{xy}\\
\sigma_{yx}& \sigma_{yy} \end {array}\right )=\\
\qquad{}\\
\qquad{}\left(\begin {array}{cc} [\nu_{i}(F_{I})]^{-1} & n\epsilon
\\ -n\epsilon & [\nu_{i}(F_{I})\nu_{a}(F_{a})]^{-1} \end
{array}\right ).
\end{array}
\end {equation}

From {Eqs.} (\ref{F11}) and (\ref{F12}) we see that the
off-diagonal components of the $\hat{\rho}$ and $\hat{\sigma}$
tensors satisfy the Onsager relation ($\rho_{xy}=-\rho_{yx}$ in
the general nonlinear case and $\sigma_{xy}=-\sigma_{yx}$). All
the components of the $\hat{\rho}$-tensor and the diagonal
components of the $\hat{\sigma}$-tensor are functions of the
current density $j$ through the external force value $F_{L}$, the
temperature $T$, the angle $\alpha$, and the dimensionless Hall
parameter $\epsilon$. For the following (see item E.2 of Sec. II)
it is important, however, to stress that the off-diagonal
components of the $\hat{\sigma}$ (i.e. the dimensional Hall
conductivity terms $\delta=n\epsilon/\rho_f$) are not influenced
by a presence of the $i$- and $a$-pins$^{16}$.

The experimentally measurable resistive responses refer to a
coordinate system tied to the current (see Fig.~1). The
longitudinal and transverse (with  respect to the current
direction) components of the electric field, $E_{\parallel}$ and
$E_{\perp}$, are related to $ E_{x}$ and $E_{y}$ by the simple
expressions
\begin{equation}
\label{F13}
\begin{array}{ll}
 E_{\parallel}=E_{x}\sin\alpha+E_{y} \cos \alpha,\\
 \\
 E_{\perp}=-E_{x}\cos\alpha+E_{y}\sin\alpha.\\
\end{array}
\end{equation}

Then according to {Eqs.} (\ref{F13}), the expressions for the
experimentally observable longitudinal and transverse (with
respect to the $\mathbf{j}$-direction ) magnetoresistivities
$\rho_{\parallel}\equiv E_{\parallel}/j $ and $\rho_{\perp}\equiv
E_{\perp}/j$ have the form:
\begin{equation}
\label{F14}
\begin {array}{ll}
\rho_{\parallel}=\rho_{xx}\sin^{2}\alpha +\rho_{yy}\cos^{2}\alpha ,\\
\\
\rho_{\perp}=\rho_{yx}+(\rho_{yy}-\rho_{xx})\sin\alpha\cos\alpha.\\
\end{array}
\end{equation}
Note, however, that the magnitudes of the
$\rho_{\parallel,\perp}$, given by {Eqs.} (\ref{F14}), are, in
general, depend on the direction of the external magnetic field
$\mathbf{B}$ along $z$ axis due to the $n\epsilon$-dependence of
the $F_{I} $ and $F_{a}$ forces in arguments of the $\nu_{i}$ and
$\nu_{a}$ functions, respectively. In order to consider only
$n$-independent magnitudes of the $\rho_{\parallel}$- and
$\rho_{\perp}$-resistivities we should introduce the even(+) and
the odd($-$) with respect to magnetic field reversal
$(\rho^{\pm}\equiv (\rho(n)\pm\rho(-n))/2)$ longitudinal and
transverse dimensional magnetoresistivities, which in view of
Eqs.~(\ref {F14}) have the form:
\begin{equation}
\label{F15}
\begin {array}{ll}
\rho_{\parallel}^{+}=\rho_{f}[\sin^{2}\alpha + \nu_{a}(F_{Lx})\cos^{2}\alpha ]\nu_{i}(F_{i}),\\
\\
\rho_{\parallel}^{-}=\rho_{f}[[\sin^{2}\alpha+\nu_{a}(F_{Lx})\cos^{2}\alpha]\nu_{i}^{-}(F_{I})+\\
\\
\qquad{} \nu_{i}(F_{i})\nu_{a}^{-}(F_{a})\cos^{2}\alpha].\\
\end{array}
\end{equation}
\begin{equation}
\label{F16}
\begin {array}{ll}
\rho_{\perp}^{+}=-\rho_{f}\nu_{i}(F_{i})[1-\nu_{a}(F_{Lx})]\sin\alpha\cos\alpha,\\
\\
\rho_{\perp}^{-}=\rho_{f}[n\epsilon\nu_{a}(F_{Lx})\nu_{i}^{2}(F_{i})+\\
\\
\qquad\{\nu_{a}^{-}(F_{a})\nu_{i}(F_{i})-\nu_{i}^{-}(F_{I})[1-\nu_{a}(F_{Lx})]\}\times\\
\\
\qquad\qquad\qquad\qquad\qquad\qquad\qquad\qquad\sin\alpha\cos\alpha].\\
\end{array}
\end{equation}

Here $\nu^{-}$ are the odd
($\nu^{-}\equiv\nu^{-}(n)=(\nu(n)-\nu(-n))/2$) components of the
functions $\nu_{i}(F_{I})$ and $\nu_{a}(F_{a})$, and for
$\nu^{-}_{a}(F_{a})$ we have the expansion in terms of
$\epsilon<<1$:
\begin{equation}
\label{F17}
 \nu_{a}^{-}\simeq
 -n\epsilon\nu_{i}(F_{i})F_{Ly}[d\nu_{a}(F_{Lx})/dF_{Lx}].
\end{equation}
{Eqs.} (\ref{F15})-(\ref{F16}) are accurate to the first order in
$\epsilon<<1$ and contain a lot of new physical information, which
will be analyzed below (see item E of Sec. II). However, before
this analysis it is instructive to discuss in short the main
physically important features of two main limiting cases of purely
anisotropic $a$-pinning and isotropic $i$-pinning, which follow
from {Eqs.} (\ref{F15})-(\ref{F16}), when $\nu_{i}=1$ or $
\nu_{a}=1$, respectively.

\subsection{Anisotropic a-pinning.}

Setting $\nu_{i}=1$ we obtain rather simple formulas, which were
derived firstly$^{11}$ for the experimentally observable nonlinear
even and odd longitudinal  and transverse anisotropic
magnetoresistivities
$\rho_{\parallel,\perp}^{\pm}(j,\theta,\alpha,\varepsilon_{a})$ as
functions of the transport current density $j$, dimensionless
temperature $\theta$ and relative volume fraction
$0\leq\varepsilon_{a}\leq1$, occupied by the parallel twin planes,
directed at an angle $\alpha$ with respect to the current
direction:
\begin{equation}
\label{F18} \rho_{\parallel
a}^{+}=\rho_{f}[\nu_{a}^{+}\cos^{2}\alpha+\sin^{2}\alpha],\quad
\rho_{\perp a}^{+}=\rho_{f}(\nu_{a}^{+}-1)\sin\alpha\cos\alpha,
\end{equation}
\begin{equation}
\label{F19} \rho_{\parallel
a}^{-}=\rho_{f}\nu_{a}^{-}\cos^{2}\alpha,\quad\rho^{-}_{\perp
a}=\rho_{f}[n\epsilon\nu_{a}^{+}+\nu_{a}^{-}\sin\alpha\cos\alpha].
\end{equation}

Here $\nu_{a}=\nu_{a}(F)$ is considered as the probability of
overcoming the potential barrier of the washboard PPP in the
$x$-direction under the influence of the effective force $F\equiv
F_{Lx}-n\epsilon F_{Ly}$$^{11}$. This $\nu_a$-function describes
an essentially nonlinear transition from the linear
low-temperature thermoactivated flux flow (TAFF) regime of vortex
motion to the ohmic flux flow (FF) regime. It is a step-like
function of $j$ or $\theta$ for a small fixed temperature or
current density respectively (see Figs. 4, 5 in Ref. 11).

It follows from Eqs. (18)-(19) that for $\alpha\not=0, \pi/2$  the
observed resistive response contains not only the ordinary
longitudinal $\rho_{||a}^+(\alpha)$ and transverse $\rho_{\perp
a}^-(\alpha)$ magnetoresistivities, but also two new components
induced by the pinning anisotropy: an {\it even transverse}
$\rho_{\perp a}^+(\alpha)$ and an {\it odd longitudinal} component
$\rho_{||a}^-(\alpha)$. The physical origin of the $\rho_{\perp
a}^+(\alpha)$ (which is independent of $\epsilon$) is related in
an obvious way with the guided vortex motion along the "channels''
of the washboard pinning potential in the TAFF regime. On the
other hand, the component $\rho_{||a}^-(\alpha)$ is proportional
to the odd component $\nu_a^-$, which is zero at $\epsilon=0$ and
has a maximum in the region of the nonlinear transition from the
TAFF to the FF regime at $\epsilon\not=0$ (see Figs. 6, 7 in Ref.
11) The $(j,\theta)$-dependence of the odd transverse (Hall)
resistivity $\rho_{\perp a}^-(j,\theta)$ has contributions both,
from the even $\nu_a^+\approx\nu_a$ and from the odd $\nu_a^-$
components of the $\nu_a(j,\theta)$-function. Their relative
magnitudes are determined by the angle $\alpha$ and the effective
Hall constant $\epsilon$. Note, that as the odd longitudinal
$\rho_{\parallel a}^-$  and odd transverse $\rho_{\perp a}^-$
magnetoresistivities arise by virtue of the Hall effect, their
characteristic scale is proportional to $\epsilon<<1$ (see Eqs.
(19)).

The appearance of these new odd Hall contributions follows from
emergence of a certain equivalence of $xy$-directions for the
case, where a guiding of the vortex along the channels  of the
washboard anisotropic pinning potential is realized$^{18}$ at
$\alpha\neq0,\pi/2$ and leads to the new specific angle-dependent
"scaling" relations for the Hall conductivity$^{11}$ for the case
$\epsilon\tan\alpha<<1$
\begin{equation}
\label{F20}n\epsilon=(\rho_{\perp a}^{-}-\rho_{\parallel
a}^{-}\tan\alpha)\cos^{2}\alpha/(\rho_{\parallel
a}^{+}-\rho_{f}\sin^{2}\alpha).
\end{equation}
Here the dimensionless Hall constant $\epsilon<<1$ is uniquely
related to three experimentally observable nonlinear resistivities
$\rho_{\parallel a}^{+}, \rho_{\parallel a}^{-}, \rho_{\perp
a}^{-}$, and the "scaling" relation (\ref {F20}) depends on the
angle $\alpha$. This relation differs substantially from the
power-law scaling  relations, obtained in the isotropic
case$^{17}$ (see below). In the particular case $\alpha=0$ we
regain the results$^{10}$, specifically $\epsilon=\rho_{\perp
a}^{-}/\rho_{\parallel a}^{+}$ (in Ref. 10
$\epsilon=\rho_{\perp}/\rho_{\parallel}$), i.e. a linear
relationship between $\rho_{\perp a}^{-}$ and $\rho_{\parallel
a}^{+}$.

Eq. (20) may be represented in another form
\begin{equation}
\rho_{\perp a}^-(\alpha)={\delta}\nu_a(\alpha){\rho_f}^2
-\rho_{||a}^-(\alpha)\tan\alpha
\end{equation}
which is more suitable for considering scaling relations in
longitudinal ($\alpha=\pi/2$) and transverse ($\alpha=0$)
LT-geometries of experiment$^{11}$. In these geometries second
term in the right hand side of Eq. (21) is zero and we obtain that
\begin{equation}
\rho_{\perp a}^-= \tilde{\delta}({\rho_{||a}^+})^2,
\end{equation}
\begin{equation}
\begin{array}{ll}
\tilde{\delta}(\alpha=\pi/2)\equiv\tilde{\delta_L}={\delta}\nu_a(0,\theta);\\
\\\tilde{\delta}(\alpha=0)\equiv\tilde{\delta_T}={\delta}/\nu_a(j,\theta).
\end{array}
\end{equation}

From Eqs. (22)-(23) follows that $\tilde{\delta}$ may be
interpreted as an \emph{effective} Hall conductivity in
LT-geometries which is suppressed for $\alpha=\pi/2$
($\tilde{\delta_L}<{\delta}$) and enhanced for $\alpha=0$
($\tilde{\delta_T}>{\delta}$) in comparison with a bare Hall
conductivity ${\delta}$. The physical reason for this influence of
$\nu_a$-function on the $\tilde{\delta}$ behavior in LT-geometries
is simple. Namely, it appears as a result of the fact that in the
case of anisotropic pinning the driving force $F\equiv
F_{Lx}-n\epsilon F_{Ly}$, which determines the probability of
overcoming the potential barrier ( and therewith also determines
the magnitude of the component of the vortex velocity
perpendicular to the channels of the PPP), is the sum of two
forces. The first of these is the transverse component of the
Lorentz force, $F_{Lx}$= $F_{L}\cos\alpha$, and the other is the
\emph{transverse} component of the Hall force $ F_{H}=n\epsilon
F_{Ly}$ which is proportional to the longitudinal (relative to the
PPP planes) component of the velocity of guided vortex motion.
This second force $ F_{H}$, which changes its sign (relative to
the sign of $F_{L}$ ) upon reversal of the sign of the external
magnetic field, is the reason for appearance of new, Hall-like in
their origin, $\nu^-$-terms in the formulas for the resistive
responses in Eqs. (19).

Returning to the physics of suppression and enhancement of the
$\tilde{\delta}$ in LT-geometries we should keep in mind that only
longitudinal component of the vortex velocity (with respect to the
current direction) $v_l$ is responsible for the appearance of the
transverse Hall voltage. Thus, in L-geometry $v_l$ and
$\tilde{\delta}$ are suppressed by PPP-barriers, whereas in
T-geometry $v_l$ is not influenced by them and $\tilde{\delta}$
looks like enhanced quantity. On the contrary, the behavior of the
transverse component of the vortex velocity $v_t$, which
determines the longitudinal voltage, in LT-geometries is opposite.

\subsection{Isotropic i-pinning.}

For this case we put $ \nu_{a}=1$ and from
{Eqs.}~(\ref{F15})-(\ref{F16}) follows that
\begin{equation}
\label{F21} \rho_{\parallel}^{+}=\rho_{\parallel i
}=\rho_{f}\nu_{i}(F_{L}),\quad \rho_{\perp}^{-}=\rho_{\perp
i}=\rho_{f}n\epsilon\nu_{i}^{2}(F_{L}),
\end{equation}
where $F_{L}=F_{i}(\nu_{a}=1)=\mid\mathbf{F}_{L}\mid$. From {Eqs.}
(\ref{F21}) the well-known scaling relation $ \rho_{\parallel
i}\sim(\rho_{\perp i})^{2}$, derived firstly in Ref.~17, follows.
Note that $\rho_{\perp i}^{+}=\rho_{\parallel i}^{-}=0$ in this
case, i.e. nonlinear resistive response is isotropic.

\subsection{Competition between a- and i-pinning.}

Equations (\ref{F15})-(\ref{F16}) for the magnetoresistivities
$\rho_{\parallel,\perp}^{\pm}$ at arbitrary value of competition
between point-like  and anisotropic planar disorder for the
in-plane geometry of experiment can be represented in a more
suitable form, if we take into account {Eqs.} (\ref {F18})-(\ref
{F19}) and (\ref{F21}):
\begin{equation}
\label{F22}\rho_{\parallel}^{+}=\nu_{i}(F_{i})\cdot\rho_{\parallel
a}^{+},\quad \rho_{\perp}^{+}=\nu_{i}(F_{i})\cdot\rho_{\perp
a}^{+},
 \end{equation}
\begin{equation}
\label{F23}\rho_{\parallel}^{-}=\nu_{i}^{-}\rho_{\parallel
a}^{+}+\nu_{i}(F_{i})\cdot\rho_{\parallel a}^{-},
 \end{equation}
\begin{equation}
\label{F24}
\rho_{\perp}^{-}=\rho_{f}n\epsilon\nu_{a}\nu_{i}^{2}+\rho_{f}\{\nu_{a}^{-}\nu_{i}-\nu^{-}_{i}[1-\nu_{a}]\}\sin2\alpha/2.
 \end{equation}

Here $\nu_{i}(F_{i})$ is the probability function $\nu_{i}$ of
anisotropic argument
$F_{i}=[F^{2}_{Lx}\nu^{2}_{a}(F_{Lx})+F_{Ly}^{2}]^{1/2}$, the
magnetoresistivity $\rho_{\parallel,\perp a}^{\pm}$ and the
$\nu_{a}\equiv\nu_{a}(F_{Lx})$-functions in
{Eqs.}~(\ref{F22})-(\ref{F24}) are the same as those in item C of
Sec. II; $\nu_{i}^{-}=\nu_{i}^{-}[F_{I}(n)]$ and
$F_{I}(n)=[F_{Ly}^{2}+
F_{Lx}^{2}\nu_{a}^{2}(F_{a})+2n\epsilon\nu_{i}(F_{i})F_{Lx}F_{Ly}\nu_{a}(1-\nu_{a})]^{1/2}$.
It is easy to check, that previous results of items C and D of
Sec. II follow from {Eqs.} (\ref{F22})-(\ref{F24}) in the limits
of purely anisotropic (i.e. for  $\nu_{i}=1$, $\nu_{i}^{-}=0$) and
isotropic (i.e. for $\nu_{a}=1$, $\nu_{a}^{-}=0$) pins.

In this subsection it must be suffice to discuss in short the main
physically important features  of these equations. First of all,
the magnetoresistivities $\rho_{\parallel,\perp}^{\pm}$ can be
found, if the $\nu_{a}$- and $\nu_{i}$- functions are known.
Moreover, the converse statement is also valid: it is possible to
reconstruct these functions from ($j$, $\theta$, $B$)-dependent
resistive measurements, using only {Eqs.} (\ref{F22}), where the
Hall terms are ignored. {Eqs.} (\ref{F23}) and (\ref{F24}), which
arise due to the Hall effect, have a rather complicated structure,
which reflects a more pronounced competition between isotropic and
anisotropic disorder in the Hall-mediated  resistive responses.
Let us outline the main new physical results, following from
{Eqs.} (\ref{F22})-(\ref{F24}).

\subsubsection{Point-like disorder and vortex
guiding.}

For the discussion of the influence of point-like pins on the
guiding of vortices in the anisotropic pinning potential it is
sufficient to analyze {Eqs.}~(\ref{F22}). Whereas for the purely
anisotropic pinning ($\nu_{i}=1$) a critical current density
$j_{c}$ exists only for direction, which is strictly perpendicular
to the PPP ($\alpha=0$) and $j_{c}(\alpha)=0$ for any other
direction ($0<\alpha\leq\pi/2$) due to the guiding  of vortices
along the channels of a washboard potential, in {Eqs.}~(\ref{F22})
the factor $\nu_{i}(F_{i})$ ensures  that an anisotropic critical
current density $j_{c}(\alpha,\theta)$ exists for arbitrary
angles~$\alpha$.

It is interesting, however, to note, that the angular dependence
of the ratio $\rho_{\perp}/\rho_{\parallel}$, which determines the
angle $\beta$ between $\mathbf{j}$ and $\mathbf{v}$ for $a$-pins
in Ref. 11, according to the relation
\begin{equation}
\label{F25}\cot\beta=-\frac{\rho_{\perp a}^{+}}{\rho_{\parallel
a}}=\frac{1-\nu_{a}}{\tan\alpha+\nu_{a}\cot\alpha}
 \end{equation}
is not influenced by the isotropic disorder, because factor
$\nu_{i}(F_{i})$ in  {Eqs.} (\ref{F22}) vanishes from {Eq.}
(\ref{F25}). Physically it means, that character of anisotropy in
the case of competition between $i$- and $a$-pinning is determined
only by
$<\mathbf{F}_{p}^{a}>=[\nu_{a}(F_{Lx},\theta)-1]\mathbf{F}_{Lx}$,
(see Fig.~1), i.e. by the average pinning force of the PPP.
Isotropic pins influence only the magnitude of the average
$\mathbf{v}$-vector, because $<\mathbf{F}_{p}^{i}> \parallel
\mathbf{v}\parallel\mathbf{F}_{I}$. So, the polar resistivity
diagram $\rho(\alpha)$, which can be measured experimentally
$^{5}$, is influenced by point-like pins, because from {Eqs.}
(\ref{F11}) follows, that
\begin{equation}
\begin {array}{ll}
\label{F26}\rho(\alpha)=\rho_{f}[\rho_{xx}^{2}\sin^{2}\alpha+\rho_{yy}^{2}\cos^{2}\alpha]^{1/2}=\\
\\
\qquad{} \rho_{f}\nu_{i}(F_{i})(\sin^{2}\alpha+\nu_{a}^{2}\cos^{2}\alpha)^{1/2}.\\
\end{array}
\end{equation}

\subsubsection{New Hall voltages and scaling relations.}

As it follows from {Eqs.} (\ref{F23})-(\ref{F24}), the odd
longitudinal $\rho_\|^-$ and transverse $\rho_\perp^-$
magnetoresistivities contain terms with the $\nu_i^-$-function.
They possess a highly anisotropic current- and
temperature-dependent bump-like behavior. They tend to zero in the
linear regime of vortex motion. For $\alpha=0,\pi/2$ these new
terms disappear, because $\nu_i^-=\nu_a^-=0$ at these limits. As
it was in the case of purely a-pinning (see item C of Sec. II),
the appearance of these new odd Hall contributions follows from
the emergence of a certain equivalence of $xy$-directions due to a
guiding of vortices along the channels of the washboard pinning
potential for the case with $\alpha\not=0,\pi/2$. Note also, that
$\rho_\|^-$ includes two terms with similar signs, whereas in
$\rho_\perp^-$ there are terms with opposite signs. The latter can
give rise to the well-known sign change in the
$(j,\theta,H)$-dependence of the Hall resistivity below $\mathrm
T_c$$^{12}$.

From {Eqs.} (\ref{F22})-(\ref{F24}) new anisotropic "scaling"
relations for the dimensionless Hall constant $\epsilon$ can be
derived. For this purpose we exclude $\nu_{i}^{-}$ from {Eqs.}
(\ref{F22})-(\ref{F24}), for $\nu_{a}^{-}$ use {Eq.}~(\ref{F17});
and after some algebra in the limit $\epsilon\cdot\tan\alpha<<1$
we have:
\begin{equation}
\label{F27}n\epsilon=\frac{2\rho_{\perp}^{-}\cdot\rho_{\parallel}^{+}+\rho_{f}\sin2\alpha(1-\nu_{a})\nu_{i}\rho_{\parallel}^{-}}{[2\nu_{a}\rho_{\parallel}^{+}-\sin2\alpha\cdot\rho_{f}\nu_{i}F_{Ly}\nu_{a}^{'}]\rho_{f}\nu_{i}^{2}}.
\end{equation}

It is easy to check that from {Eq.}~(\ref{F27}) follows scaling
relations $\delta=n\epsilon/\rho_{f}=\rho_{\perp
i}/(\rho_{\parallel i})^2$ (for $i$-pins at $\nu_a=1$) and
{Eq.}~(\ref{F20}) (for $a$-pins at $\nu_i=1$)

As it follows from Eqs. (25) and (27), just the same "scaling"
relations as given by Eqs. (22) and (23) for \emph{a}-pins, exist
also for (\emph{i}+\emph{a})-pins (with a replacement of
corresponding \emph{a}-resistivities in Eq. (8) by
$\rho_f\rho_{\perp}^{-}$ and $\rho_f\rho_{\parallel}^{+}$).
Physically it follows from the fact that point-like disorder does
not change the angular dependence of the ratio $\rho_{\perp
a}^{+}/\rho_{\parallel a}^{+}$, which determines the angle $\beta$
between $\mathbf{j}$ and average velocity vector
$\langle\mathbf{v}\rangle$ for $a$-pins$^{11}$, and influences
only the magnitude of $\langle\mathbf{v}\rangle$$^{16}$.

\section{Grafical analysis of nonlinear  regimes.}

\subsection{Pinning potential and $\nu$-function behavior.}

 In order to analyze different types of nonlinear anisotropic $(j, \theta,
\alpha)$-dependent magnetoresistivity  responses, given by
formulas (\ref{F22})-(\ref{F24}), we should bear in mind that
these responses, as is seen from formula (\ref{F11}), are
completely determined by the $(j,\theta)$-behavior of the
functions $\nu_{a}(F_{a})$ and $\nu_{i}(F_{I})$, having a sense of
the probabilities to overcome the effective potential barriers of
the $a$- and $i$-pins, respectively. A simple analytical model for
the calculation of the $(j, \theta)$-dependent $\nu$-functions was
given earlier$^{11,13,20}$. We will use for both $\nu_{i}$ and
$\nu_{a}$ functions the one-dimensional periodic pinning potential
$U_{p}(x)$ (see Fig.~2), which has a simple analytical
form$^{11,20}$:
\begin{equation}
\label{F28} U_{p}(x)= \left\{
 \begin{array}{crr}
 \, - F_{p}x, \qquad\qquad  0 \leqslant x \leqslant b,\\
\\
 F_{p}(x-2b), \qquad  b  \leqslant x \leqslant 2b,\\
 \\
  0, \qquad\qquad\qquad  2b \leqslant x \leqslant h , \\
\end{array}
\right.
 \end{equation}
where $F_{p}$ is the pinning force ($F_{p}=U_{0}/b$, where
$U_{0}>0$ is the depth of the potential well and $2b$ is the width
of the well). This form of $U_{p}(x)$ allows to define as the
properties of a given pinning center (by the parameters $U_{0}$
and $b$), as well as the density of such centers (by the parameter
$\varepsilon=2b/h$, where $h$ is the period of the $U_{p}(x)$).

\begin{figure}
\includegraphics{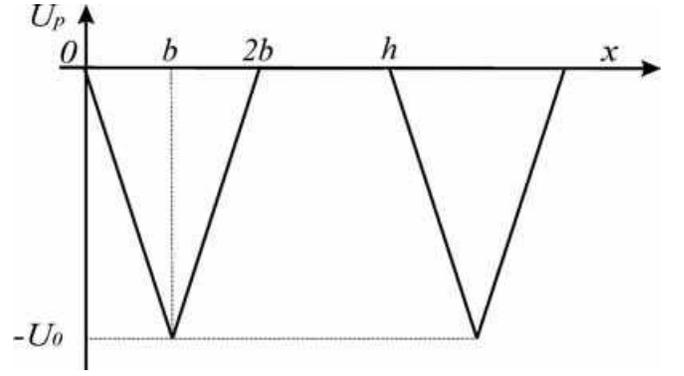}
\caption{Model pinning potential $U_{p}(x)$: $h$ is the period of
the potential, $2b$ is the width of the potential well, $U_{0}$ is
the depth of the potential well, $\varepsilon=2b/h$ characterizes
the concentration of the pinning planes.} \label{fig2}
\end{figure}

Calculation of the $\nu(j, \theta)$ function on the basis of the
pinning potential, given by {Eq.}~(\ref{F28}), was done $^{11}$
and can be represented here in the form$^{11}$
\begin{equation}
\label{F29}
 \nu(f, \theta,\varepsilon)=\frac{2f(f^{2}-1)^{2}}{2f(f^{2}-1)(f^{2}-1+\varepsilon)-\varepsilon\theta G},
\end{equation}

where
$G=\{(3f^{2}+1)\cosh(f/\theta\varepsilon)+(f^{2}-1)\cosh[(f(1-\\
\\
\qquad2\varepsilon))/(\theta\varepsilon)]-2f(f-1)\cosh[f(1-\varepsilon)/\theta\varepsilon-(1/\theta)]-\\
\\
\qquad\quad2f(f+1)\cosh[(f(1-\varepsilon)/\theta\varepsilon+1/\theta]\}/\sinh(f/\theta\varepsilon).
$\\

Here and below we have for the time being dropped the indices $a$
and $i$ from the physical quantities pertaining to pinning
potentials $U_{pa}$ and $U_{pi}$ and formula (\ref{F29}) describes
equally the pinning on both potentials. For convenience of
qualitative analysis of the formulas following dimensionless
parameters were used: $f=Fb/U_{0}$ is the effective motive force,
which specifies its ratio to the pinning force $F_{p}=U_{0}/b$,
$\theta=T/U_{0}$ is the temperature.

The effect of the external force $F$ acting on the vortices
consists in a lowering of the potential barrier for vortices
localized at pinning centers and, hence, an increase in their
probability of escape from them. Increasing the temperature also
leads to an increase in the probability to escape of the vortices
from the pinning centers through an increase in the energy of
thermal fluctuations of the vortices. Thus the pinning potential
of a pinning center, which for $F, T\rightarrow0$ leads to
localization of the vortices, can be suppressed by both an
external force and by temperature.

A detailed quantitative and qualitative analysis of the behavior
of $\nu(f,\theta,\epsilon)$ as a function of all the parameters
and its asymptotic behavior as a function of each  are
described$^{11}$. Here we will pay particular attention only to
the typical curves of $\nu(f,\theta,\epsilon)$ as a function of
the parameters $f$ and $\theta$, which describe the nonlinear
dynamics of the vortex system as a function of the external force
acting on the vortices in the direction perpendicular to the
pinning centers and as a function of temperature (see Figs. $4$
and $5$ in Ref. 11). As we see from those figures, the form of the
$\nu(f)$ and $\nu(\theta)$ curves is determined by the values of
the fixed parameters $\theta$ and $f$. The monotonically
increasing function $\nu(f)$ reflects the nonlinear transition of
the vortex motion from the TAFF to the FF regime with the
increasing external force at low temperatures ($T \ll U_{0}$),
while at high temperatures ($T \gg U_{0})$ the FF regime is
realized in the entire range of variation of the external force
(even at small forces) because of the effect of thermal
fluctuation on the vortices. The monotonically increasing function
$\nu(\theta)$ reflects the nonlinear transition from a dynamical
state corresponding to the value of the external force at zero
temperature to the FF saturation regime. The width of the
transition from the TAFF to the FF regime on the $\nu(f)$ and
$\nu(\theta)$ curves depends on substantially different on the
increasing of the parameters $\theta$ and $f$, respectively.
Namely, with increasing $\theta$ the function $\nu(f)$ shifts
leftward and becomes less steep (see Fig.~4 in Ref.~11). That is,
the higher the temperature, the smoother the transition from the
TAFF to the FF regime and the lower the values of the external
force, at which it occurs. With increasing $f$ the $\nu(\theta)$
curve also shifts leftward, it becomes steeper (see Fig.~5 in
Ref.~11). Consequently, the greater the suppression of the
potential barrier of the pinning center by the external force, the
sharper the transition from the TAFF to the FF regime and the
lower the temperature at which it occurs.

These graphs will be needed later on when we will discuss the
physical interpretation of the observed guiding-depended resistive
responses. We also note that the dependence of the probability
function $\nu(\varepsilon)$ on the concentration of pinning
centers decreases monotonically from the value $\nu(0)=1$, which
corresponds to the absence of pinning centers, and that it becomes
steeper with decreasing fixed parameters $f$ and $\theta$, owing
to the growth of the probability  density for finding the vortices
at the pinning centers with decreasing temperature and external
force.

\subsection{Dimensionless form of the
$\rho_{\parallel,\perp}^{\pm}$-responses.}

Let us turn to the dimensionless parameters by which one can in
general case take into account the difference of the potentials
$U_{a}$ and $U_{i}-$ specifically, the difference of their periods
$h_{a}$, $h_{i}$, the potential well depths $U_{0a}$, $U_{0i}$ and
the width $b_{a}$, $b_{i}$. We introduce some new parameters:
$\varepsilon=(\varepsilon_{a}\varepsilon_{i})^{1/2}$ is the
average concentration of pinning centers,
$U_{0}=(U_{0a}U_{0i})^{1/2}$ is the average depth of potential
well,
$\kappa=(\varepsilon_{i}/\varepsilon_{a})^{1/2}=(h_{a}b_{i}/h_{i}b_{a})^{1/2}$,
and $p=(U_{0a}/U_{0i})^{1/2}$ , where the parameters $\kappa$ and
$p$ are measures of the corresponding anisotropies. The
temperature will be characterized by new parameters:
$\theta_{a}=pT/U_{0}=T/U_{0a}$ and
$\theta_{i}=(1/p)T/U_{0}=T/U_{0i}$, which are the ratio of the
energy of thermal fluctuations of the vortices to the average
potential well depth $U_{0a}$ and $U_{0i}$, respectively.

The current density will be measured in units of
$j_{c}=cU_{0}/\Phi_{0}h$, where $h=(h_{a}h_{i})^{1/2}$. Then the
dimensionless parameters $f_{a}$ and $f_{i}$, which specify the
ratio of the external forces $F_{a}$ and $F_{i}$ to the pinning
forces $F_{pa}=U_{0a}/b_a$ and $F_{pi}=U_{0i}/b_i$ ($\nu_{a}$ and
$\nu_{i}$ are the even functions of their arguments), we denote as
$f_{a}=F_{a}/F_{pa}$ and $f_{i}=F_{i}/F_{pi}$. The values of the
external force $F$, at which the heights of the potential barriers
$U_{0a}$ and $U_{0i}$ vanish at $T=0$ correspond (at $\alpha=0$
and $\alpha=\pi/2$) to the critical current densities
$j_{ca}=qj_{c}$ and $j_{ci}=j_{c}/q$ respectively, where
$q=p/\kappa$. In general case of nonzero temperature and
$0<\alpha<\pi/2$ it is possible to consider the angle-dependent
crossover current densities $j_{ca}(\alpha)$ and $j_{ci}(\alpha)$
(see below) which correspond to change in the vortex dynamics from
the TAFF regime to a nonlinear regime. The condition, that
determines the temperature region, in which the concept of
critical current densities is physically meaningful is $0
\leqslant T \ll U_{0}$, because for $T \gtrsim U_{0}$ the
transition from the TAFF to the nonlinear regime is smeared, and
the concept of critical current loses its physical meaning.

It is possible now to rewrite {Eqs.} (\ref{F15})-(\ref{F16}) in
the dimensionless form in order to represent them as functions of
$j$, $\theta$, $\alpha$ at given values of parameters
$\varepsilon$, $\epsilon$, $q$, $k$.

\begin{equation}
\label{F30}\rho_{\parallel}^{+}=\nu_{i}(f_{i})[\sin^2\alpha+\nu_{a}(f_{a})\cos^2\alpha)],
\end{equation}
\begin{equation}
\label{F31}\rho_{\perp}^{+}=-\nu_{i}(f_{i})[1-\nu_{a}(f_{a})]\sin2\alpha/2,
\end{equation}
\begin{equation}
\label{F32}\rho_{\parallel}^{-}=\nu_{i}^{-}(\tilde{f_{i}})[\sin^2\alpha+\nu_{a}(f_{a})\cos^2\alpha)]+\nu_{a}^{-}(\tilde{f_{a}})\nu_{i}(f_{i})\cos^{2}\alpha,
\end{equation}\\
\begin{equation}
\label{F33}
\begin{array}{l}\rho_{\perp}^{-}=n\epsilon\nu_{a}(f_{a})\nu_{i}^{2}(f_{i})+ \\ \\ +\{\nu_{a}^{-}(\tilde{f_{a}})\nu_{i}(f_{i})-
\nu^{-}_{i}(\tilde{f_{i}})[1-\nu_{a}(f_{a})]\}\sin2\alpha/2,
\end{array}\end{equation}\\
\begin{equation}
\label{F34}
\textrm{where} \qquad f_{a}=jq^{-1}\cos\alpha,\quad{}\\
\end{equation}
\begin{equation}
\label{F35} f_{i}=jq(\sin^2\alpha+\nu_{a}^2\cos^2\alpha)^{1/2},
\end{equation}
\\
\qquad and \qquad
$\tilde{f_{a}}=jq^{-1}[\epsilon\nu_{i}(f_{i})\sin\alpha+n\cos\alpha]$,
\begin{equation*}
\begin{array}{ll}\tilde{f_{i}}=jq\{\sin^2\alpha+\nu_{a}^2(\tilde{f_{a}})\cos^2\alpha)-n\epsilon\nu_{i}(f_{i})\nu_{a}(f_{a})\times
\\
\\\qquad\qquad\qquad\qquad\qquad\qquad\times [1-\nu_{a}(f_{a})]\sin2\alpha\}^{1/2}.
\end{array}\end{equation*}
Here \\
\\ $\nu_{a}(f_{a})=\nu_{a}(f_{a}, \theta_{a},
\varepsilon_{a}/\kappa)$,\qquad
$\nu_{a}(\tilde{f_{a}})=\nu_{a}(\tilde{f_{a}}, \theta_{a},
\varepsilon_{a}/\kappa)$,\\
\quad{}
\\
\qquad\qquad$\nu_{i}(f_{i})=\nu_{i}(f_{i}, \theta_{i},
\varepsilon_{i}\kappa)$, \qquad\quad
$\nu_{i}(\tilde{f_{i}})=\nu_{i}(\tilde{f_{i}}, \theta_{i},
\varepsilon_{i}\kappa)$,\\
\begin{equation*}
\begin{array}{crr}
\textrm{and} \qquad &\nu^{\pm}_{i}(\widetilde{f}_{i})=\{\nu_{i}[\widetilde{f}_{i}(n)]\pm\nu_{i}[\widetilde{f}_{i}(-n)]\}/2,&\\
\\
\quad
&\nu^{\pm}_{a}(\widetilde{f}_{a})=\{\nu_{a}[\widetilde{f}_{a}(n)]\pm\nu_{a}[\widetilde{f}_{a}(-n)]\}/2.&
\end{array}
\end{equation*}

In {Eqs.}~(33)-(38) we also denoted
$\nu_{i}(f_{i})\equiv\nu^{+}_{i}(\widetilde{f_{i}})$ and
$\nu_{a}(f_{a})\equiv\nu^{+}_{a}(\widetilde{f_{a}})$ for
simplicity.

Before following graphical analysis of the
$\rho_{\parallel,\perp}^{\pm}$ dependences given by  {Eqs.}
(\ref{F30})-(\ref{F33}), we should point out the magnitude of some
parameters which will be used for presentation of the graphs. It
is important to remind here that the parameter $q$ determines the
value of anisotropy between $\nu_{i}$ and $\nu_{a}$ critical
current densities, whereas the parameter $k$ describes the
anisotropy magnitude of the width of nonlinear transition from the
TAFF to the FF regime for $\nu_{i}$ and $\nu_{a}$ function. More
definitely, if $q>1$, then $j_{ca}=qj_{c}>j_{ci}=j_{c}/q$ and
influence of the $i$-pins on the vortex dynamics decreases with
$q$-increasing. For $q<1$ the situation is opposite and anisotropy
effects may be fully suppressed with $q$-decreasing. So, for the
observation of pronounced competition between $i$- and $a$-pins
$q\approx1$ should be taken.

The temperature dependences of the $\rho_{\parallel}^{+}(\alpha)$
at small current densities under conditions of the presence both
isotropic and anisotropic pinning potential were studied
experimentally$^{8}$. Arrhenius analysis of these dependences
within the frames of suggested here theoretical approach have
shown that for the samples$^{8}$ the $U_{0a}=4031$K,
$U_{0i}=1568$K, $b_{a}=400$ nm, $b_{i}=2000$ nm at $T\approx8$K.
Then for these samples $q\approx1.6$, $\kappa\approx0.5$,
$\theta\approx0.003$. It was also pointed out$^{8}$ that the best
fitting of the experimental and theoretical curves was established
for $b_{i}/b_{a}=15$, from which follows $\kappa\approx0.25$. So
for all graphs below we used $q=1.6$, $\kappa=0.25$,
$\theta=0.003$, $\epsilon=0.01$ and if it is not pointed out
specially, $\varepsilon_{a}=1$ and $\varepsilon_{i}=0.1$.

Note also that for the even longitudinal resistivity
$\rho^{+}_{\parallel}$ and the even transverse resistivity
$\rho^{+}_{\perp}$ for a small Hall effect, terms proportional to
$\epsilon\ll 1$ are absent (see {Eqs.}~(\ref{F30})-(\ref{F31}))
and only contributions describing the competition between
isotropic pinning and nonlinear guiding effect on the PPP in terms
of the even $\nu_{i}$ and $\nu_{a}$ functions are presented.

\subsection{Graphical analysis of current-angular dependences.}

\subsubsection{$(j,\alpha)$-presentation of  $\nu_{a}$ and $\nu_{i}$.}

In order to discuss graphical $(j,\alpha)$-behavior of the
resistive responses we will use $\nu_{a}$ and $\nu_{i}$ functions
of their arguments $f_{a}$ and $f_{i}$, respectively, in the form
given by {Eq.}~(\ref{F29}). Then these functions are, as a
corresponding $\nu$-function$^{11}$, the step-functions in $j$ (at
fixed $\theta$) or in $\theta$ (at fixed $j$). For every of the
$\nu$-functions it is useful to determine the "crossover current
densities" $j_{ci}(\alpha)$ and $j_{ca}(\alpha)$, as those which
correspond to the middle point of a sharp step-like nonlinear
transition from the TAFF to the FF regime. As it follows from
{Eqs.}~(\ref{F34})-(\ref{F35}), we can present $f_{a}$ and $f_{i}$
as $f_{a}=j/j_{ca}(\alpha)$ with $j_{ca}(\alpha)=q/\cos\alpha$,
and $f_{i}=j/j_{ci}(\alpha)$ with $j_{ci}(\alpha)\approx
1/q\cos\alpha$ for $\alpha\ll\pi/4$ and $j_{ci}(\alpha)\approx
1/\nu_{a}q\cos\alpha$ for
$\tan^{2}\alpha\ll\nu_{a}^{2}(j,\alpha)$; $j_{ci}(\alpha)\approx
1/q\sin\alpha$ for $\alpha>\pi/4$ because {Eq.}~(\ref{F35}) can be
presented in two equivalent forms, namely
$f_{i}=jq\cos\alpha\sqrt{\tan\alpha^{2}+\nu_{a}^{2}}=
jq\sin\alpha\sqrt{1+(\nu_{a}/\tan\alpha)^{2}}$.

\begin{figure}
\includegraphics{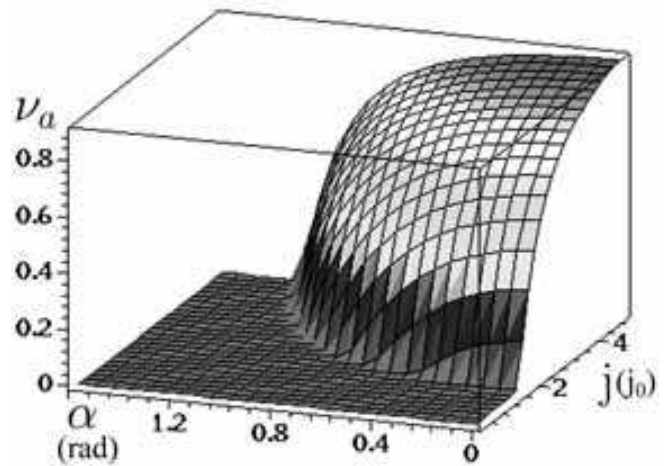}
\caption{The current-angle dependence of the anisotropic
probability function $\nu_{a}(j,\alpha)$. In all following graphs
the parameters $q=1.6$, $\kappa=0.25$, $\theta=0.003$,
$\epsilon=0.01$, $\varepsilon_{a}=1$, and $\varepsilon_{i}=0.1$
(unless otherwise stated).} \label{fig3}
\end{figure}

\begin{figure}
\includegraphics{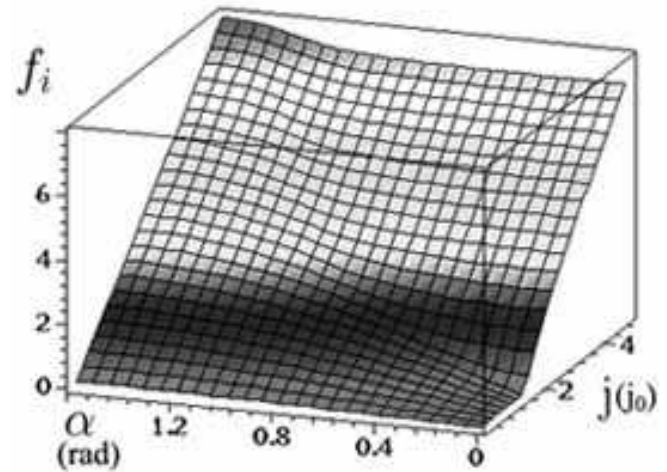}
\caption{The current-angle dependence of the average effective
motive force for a vortex $f_{i}(j,\alpha)$.} \label{fig4}
\end{figure}

\begin{figure}
\includegraphics{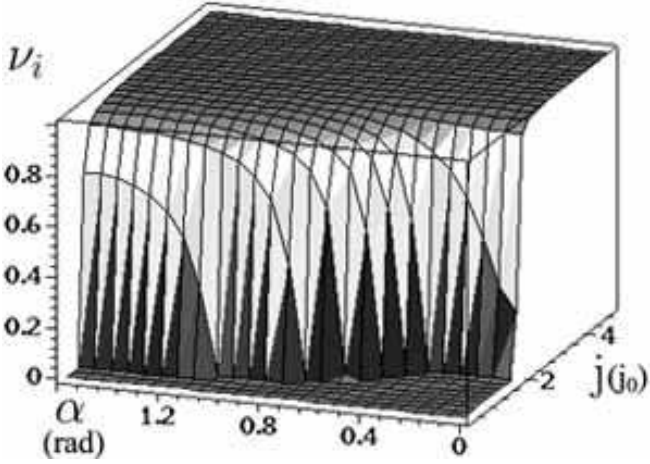}
\caption{The current-angle dependence of the isotropic probability
function $\nu_{i}(j,\alpha)$.} \label{fig5}
\end{figure}

The behavior  of  $\nu_{a}(j,\alpha)$ function (see Fig. 3) is
rather evident from  the  $f_{a}(j,\alpha)$ behavior. Namely, for
all $\alpha\neq\pi/2$ (i.e.  for $f_{a}\neq0$)) the $\nu_{a}$ with
a current increasing consistently follows next stages: a) slow
increasing at $0<j\lesssim j_{ca}$ in the TAFF regime, where
$\nu_{a}\ll1$, b)~sharp step-like increasing with a width of the
order of $j_{ca}$ which corresponds to nonlinear transition from
the TAFF to the FF regime, c)~second stage of slow increasing for
$j\gtrsim2j_{ca}$ which corresponds to the FF regime (see also
item C of Sec. II). It follows from the expression for
$j_{ca}(\alpha,q)$ that  an increasing  of $\alpha$ and~(or) $q$
leads to a broadening of the step of the  order  of $j_{ca}$ and
its shift to the larger current densities $j\approx j_{ca}$.

The anisotropy of $f_{i}(\alpha)$ (see {Eq.}~(\ref{F35}) and
Fig.~4) can be divided into two types: simple ("external") which
depends on $\cos^2\alpha$, and more complex ("internal"), given by
$\nu_{a}(\alpha)$. The first (external) anisotropy  stems from the
"tensorial" $\alpha$-dependence which exists also in the linear
(TAFF and FF) regimes of the flux motion. The second (internal) is
through the $\alpha$-dependence of $\nu_{a}$, which in the region
of transition from the TAFF  to the FF  regime is substantially
nonlinear ({Eq.}~(\ref{F29}) and Fig.~3). The appearance of
nonzero $\sin^2\alpha$ term in $f_{i}$  for $\alpha\neq0$
physically describes the guiding of vortices along the channels of
the PPP in the presence of $i$-pins for the current densities
$j\lesssim~j_{ci}(\alpha)$. The influence of $\nu_{a}$-anisotropy
on $\nu_{i}$ is different for different values of the angle
$\alpha$ (see Fig.~5). For $\alpha>\pi/4$ the anisotropy of
$\nu_{a}(\alpha)$ does not influence the value of $f_{i}(\alpha)$
because $(\nu_{a}/\tan\alpha)^{2}\ll 1$ in the expression for
$j_{ci}(\alpha)$. On the contrary, for $\alpha\ll\pi/4$ the
influence of $a$-pins on $\nu_{i}(\alpha)$ is most effective for
that range of current density, where $\nu_{a}^{2}>\tan^{2}\alpha$,
due to the inequality $\tan^{2}\alpha\ll1$. Thus, the $\nu_{i}$
and $\nu_{a}$ as functions of the angle $\alpha$ at $j=const$
behave themselves oppositely (see Figs. 3, 5): $\nu_{i}$ increases
monotonically with $\alpha$-increasing, whereas $\nu_{a}$ -
monotonically decreases. For $j\gtrsim~j_{ca}(\alpha)$ and at
small angles which meet the  condition $\tan^{2}\alpha\ll 1$, the
behavior of the $\nu_{i}$ and $\nu_{a}$ qualitatively  similar in
$\alpha$ and opposite in $q$.

In case where $\tan^{2}\alpha>1$, the $\nu_{i}$ and $\nu_{a}$
behavior is qualitatively different and stems from the
($\alpha,q$)-dependences of the corresponding crossover current
densities. In contradistinction to $\nu_{a}$, the transition of
$\nu_{i}$ from the TAFF to the FF depends weakly from $\alpha$ and
$q$; it moves to the lower current densities with $q$-increasing
for $\alpha>\pi/4$ and moves to the higher ones for
$\alpha\ll\pi/4$. In general, the $\nu_{a}$ behavior is more
anisotropic than $\nu_{i}$ behavior. The $\nu_{i}$ anisotropy
appears only in the TAFF regime, whereas $\nu_{a}$ anisotropy
exists as in the TAFF, as well in the FF regime. And this
anisotropy is greater in the current density as the angle $\alpha$
is greater. The $\nu_{i}$ and $\nu_{a}$ transition width at
$\alpha=const$ is defined by $\varepsilon_{i}$ and
$\varepsilon_{a}$ parameters, respectively, and it increases for
$\varepsilon_{i}\rightarrow 1$ and $\varepsilon_{a}\rightarrow 1$.

\subsubsection{$(j,\alpha)$-presentation of  even magnetoresistivities.}

Now we are in a position to discuss the results of the
presentation of {Eq.}~(\ref{F22}) in the form of graphs. First we
note that according to {Eqs.}~(\ref{F22}), the even resistive
responses can be represented as the products of corresponding
isotropic and anisotropic $\nu$-functions. For this reason the
graphical analysis of the $\rho_{\parallel}^{+}(j,\alpha)$ and
$\rho_{\perp}^{+}(j,\alpha)$, after the above-mentioned
consideration of the $\nu_{i}(j,\alpha)$ (see Fig.~5), can be
reduced to the construction and analysis of the $\rho_{\parallel
a}^{+}(j,\alpha)$ and $\rho_{\perp a}^{+}(j,\alpha)$ graphs.

\begin{figure}
\includegraphics{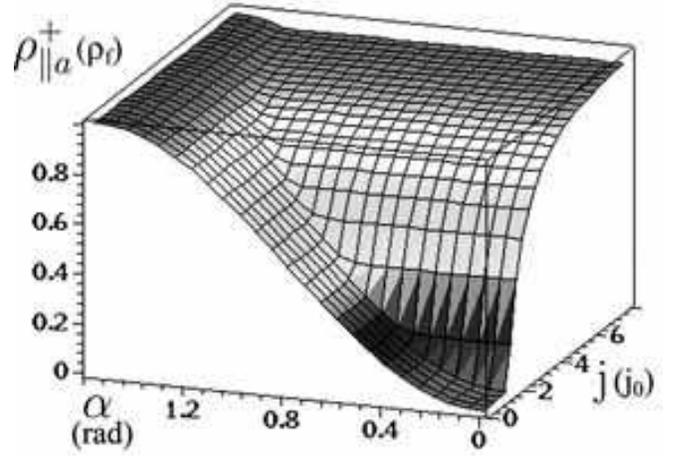}
\caption{The current-angle dependence of the dimensionless even
longitudinal anisotropic magnetoresistivity $\rho_{\parallel
a}^{+}(j,\alpha)$ for the value of the parameter
$\varepsilon_{a}=1$.} \label{fig6}
\end{figure}

\begin{figure}
\includegraphics{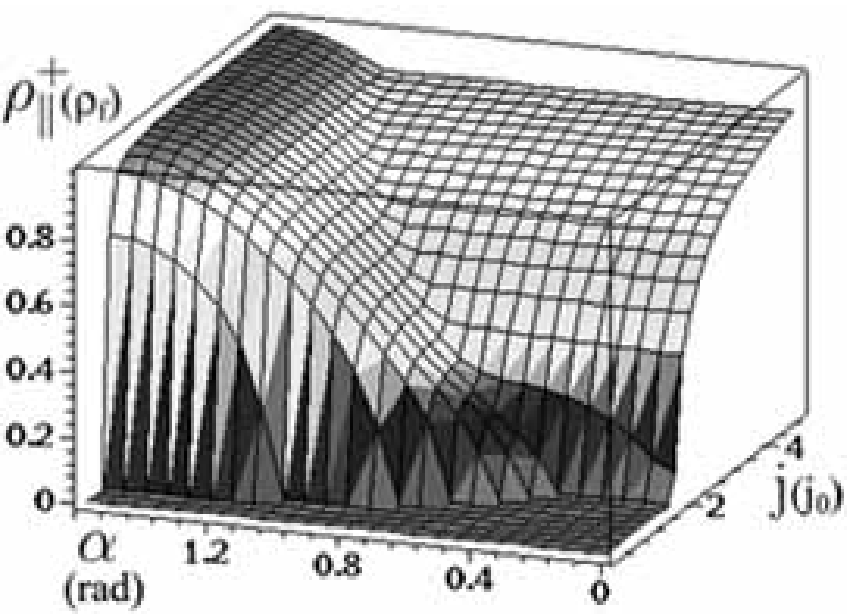}
\caption{The current-angle dependence of the dimensionless even
longitudinal magnetoresistivity $\rho_{\parallel}^{+}(j,\alpha)$
for the value of the parameter $\varepsilon_{a}=1$.} \label{fig7}
\end{figure}

\begin{figure}
\includegraphics{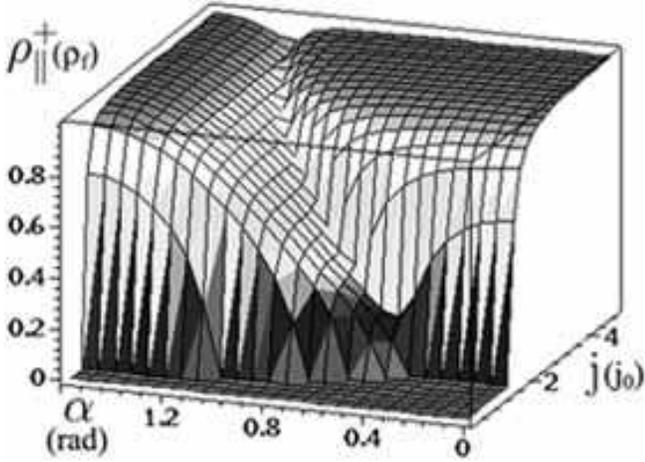}
\caption{The current-angle dependence of the dimensionless even
longitudinal magnetoresistivity $\rho_{\parallel}^{+}(j,\alpha)$
for the value of the parameter $\varepsilon_{a}=0.01$.}
\label{fig8}
\end{figure}

Let us begin with a discussion of $\rho_{\parallel a }^{+}$
behavior (see {Eq.}~(\ref{F18}) and Fig.~6). For all $\alpha\neq
0$ due to the term $\sin^2\alpha$ in {Eq.}~(\ref{F18}) a critical
current density $j_{c}$ exists only for direction, which is
strictly perpendicular to the PPP ($\alpha=0$) (as it was shown in
item E.1 of Sec. II) and $j_{c}(\alpha)=0$ for any other direction
($0<\alpha\leq\pi/2$) due to the guiding of vortices along the
channels of a washboard potential (see also Fig.~8 in Ref.~11). In
the FF-regime the isotropization of the $\rho_{\parallel a}^{+}$
arises due to the vortex slipping over the PPP channels. Thus at
small angles $\alpha$ the $\nu_{a}$ function strongly influences
the $\rho_{\parallel a }^{+}$, whereas for $\alpha\rightarrow
\pi/2$ this influence is not so effective due to the external
anisotropy, which is proportional to the $\sin^2\alpha$ term.

Returning now to the consideration of the
$\rho_{\parallel}^{+}(j,\alpha)$ graph we refer to the
{Eq.}~(\ref{F30}). It is necessary to pay special attention to the
TAFF behavior of these curves at small currents and temperatures,
which follows from the full pinning of vortices by point-like
pins. This behavior is completely different (for $\alpha\neq0$)
from the non-TAFF behavior of the corresponding graphs for the
case of purely anisotropic pinning (see Fig. 8 in Ref.~11), which
is provocated by the guiding  of vortices along the channels of
the PPP. At high current densities and (or) temperatures appears
the FF regime, because the vortex motion transverse to the
$a$-pins becomes substantial and longitudinal resistivity
practically becomes  isotropic. In these limiting cases the
$\rho^{+}_{\parallel}(j)$ magnitudes are equal to unity (Fig. 7).

For the angles $0<\alpha<\pi/2$ the $\rho_{\parallel}^{+}(j)$
behavior follows substantially the properties of one multiplier.
The qualitative behavior of these multipliers, depending on the
$j$ and $\alpha$ magnitude is very different as determined by
different behavior of their crossover current densities $j_{ci}$
and $j_{ca}$. The priority of a sharp rise of the appearance
$\nu_{i}$ or $\nu_{a}$ functions depends on the competition
between the crossover current densities $j_{ci}$ and $j_{ca}$,
respectively. That is why it may appear a "step" on some of the
$\rho_{\parallel}^{+}(j)$ curves (for $q>1$ and $\alpha\neq 0,
\pi/2$) when the next sequence of the vortex motion regimes is
realized: a)~full $i$-pinning in the TAFF regime
($0<j\lesssim~j_{ci}$); b)~nonlinear transition from the TAFF to
the FF regime for $i$-pins ($j\gtrsim2j_{ci}$), c)~practically
linear the FF regime as a consequence of the guiding of vortices
along the channels of the washboard PPP (on the
$\rho_{\parallel}^{+}(j,\alpha)$ surface one can see the
horizontal sections at $j\approx j_{ca}$, see Figs.~7,~8);
d)~nonlinear transition to the FF regime of vortex motion
transverse to the $a$-pins for $j\gtrsim j_{ca}$ and, at last,
e)~a free FF motion for $j\gg j_{ca}$.

With decreasing of the $q$ the a)-e) corresponding regions along
the current density axis $j$ can overlap each other and a common
nonlinear transition appears instead of b)-d) regions. For the
limiting cases $\alpha=0,~\pi/2$, a guiding of vortices is absent
and the $\rho_{\parallel}^{+}(j)$ LT-behavior is simply related to
the $\nu_{i}$ and $\nu_{a}$ behavior. If parameter
$\varepsilon_{a}$ is decreasing, then the width of the transition
of $\nu_{a}$ from the TAFF to the FF is also decreasing. Such
enhancement of the $\nu_{a}$ steepness leads to appearance of the
minimum in $\alpha$ for the $\rho_{\parallel}^{+}(j,\alpha)$ graph
(see Fig. 8).

\begin{figure}
\includegraphics{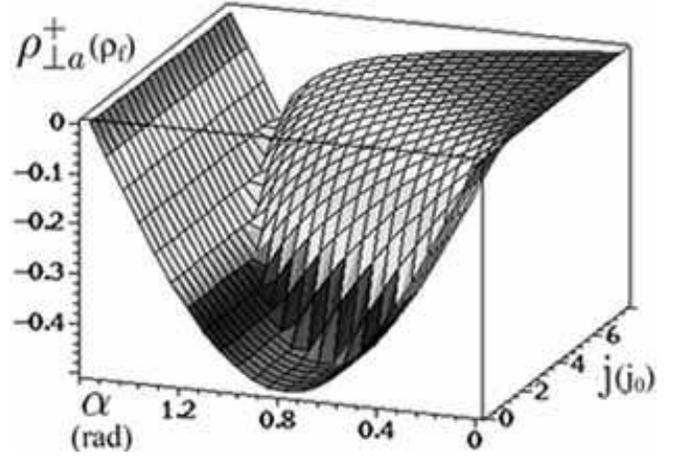}
\caption{The current-angle dependence of the dimensionless even
transverse anisotropic magnetoresistivity $\rho_{\perp
a}^{+}(j,\alpha)$} \label{fig9}
\end{figure}

\begin{figure}
\includegraphics{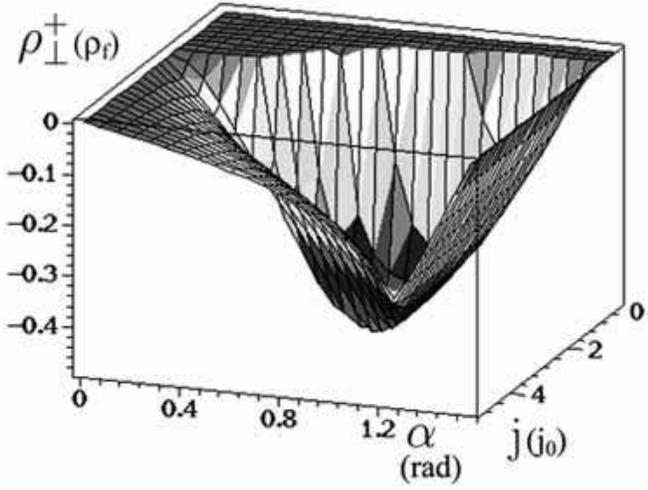}
\caption{The current-angle dependence of the dimensionless even
transverse magnetoresistivity $\rho_{\perp}^{+}(j,\alpha)$. Pay
attention to the inverted direction of the axes in comparison with
Fig.9.} \label{fig10}
\end{figure}

Now we pass to a discussion of the $\rho_{\perp a}^{+}(j,\alpha)$
and $\rho_{\perp}^{+}(j,\alpha)$ graphs. As it follows from
{Eq.}~(\ref{F18}), the $\rho_{\perp a}^{+}<0$ and has a minimum in
$\alpha$ for all $j=const$. The $\rho_{\perp a}^{+}$ reaches its
maximal magnitude for $\alpha\approx \pi/4$ due to the factor
$\sin\alpha\cos\alpha$ and realization of guiding in the TAFF
regime for $j\lesssim~j_{ca}$ (see. Fig. 9a in Ref.~11 and Fig.9).
Therefore, the most favorable angle for its observation is near
$\alpha=\pi/4$. In considered case the origin of this minimum has
the same reason as a low ($j$, $\alpha$)-behavior of the
$\rho^{+}_{\parallel}(j,\alpha)$ curves in Fig. 7, namely it stems
from existence of the TAFF regime for the point-like pins at small
$j$-values. As is seen in Fig.~9, the position and the magnitude
of this $\rho^{+}_{\perp a}$-minimum strongly depends on the
$\alpha$-value. It is very much pronounced for $q>1$ and strongly
suppressed for $q<1$ by influence of the $i$-pins. With increasing
of the current density $j\gtrsim~j_{ca}$ a position of the minimum
in $\alpha$ is shifting due to the competition of two multipliers
in the $\rho_{\perp a}^{+}(j,\alpha)$ expression
{Eq.}~(\ref{F18})): $\sin2\alpha$ is decreasing for
$\alpha\rightarrow\pi/2$, whereas $(1-\nu_{a}(j,\alpha))$ is
increasing with $\alpha$-increasing for $j=const$, and decreasing
for $j$-increasing for $\alpha=const$ due to the transition to the
FF regime. For all $\alpha$ and current densities
$j\gtrsim2j_{ca}$ the $\nu_{a}\approx1$, and for this reason
$\rho_{\perp a}^{+}\rightarrow0$. The $q$-influence is defined by
$j_{ca}(\alpha)$ and determines the region of appearance of a
small value of the $\rho_{\perp a}^{+}$ for the current densities
$j\approx j_{ca}$.

Since the $\rho_{\perp}^{+}$, according to {Eq.}~(\ref{F31}), is
the product of the $\rho_{\perp a}^{+}$ and $\nu_{i}(f_{i})$, so
this graph (see Fig. 10) can be reduced to the product of the
graphs in Fig.~5 and Fig.~9. The transition from the TAFF to the
FF regime is highly anisotropic in $\alpha$; this causes a shift
of the maximal $\rho_{\perp}^{+}(j,\alpha)$ magnitude in the
direction of a small angle $\alpha\ll\pi/4$ for the $j=const$.
That is why in view of $i$-pinning presence the
$\rho_{\perp}^{+}(j,\alpha)$, as distinct from $\rho_{\perp
a}^{+}(j,\alpha)$, has the minimum both in $\alpha$ and in $j$.
This statement follows from the fact that influence of $i$-pinning
leads to the $\rho_{\perp}^{+}\rightarrow0$ for
$0<j\lesssim~j_{ci}(\alpha)$ due to the $\nu_{i}\ll1$. For the
current densities $j\gtrsim~j_{ca}(\alpha)$ the
$\rho_{\perp}^{+}(j,\alpha)$ behavior is determined exclusively by
the above-mentioned $\rho_{\perp a}^{+}(j,\alpha)$ behavior.

\subsubsection{$(j,\alpha)$-presentation of odd
magnetoresistivities.}

Before following discussion of the odd resistive responses we
should remind the reader about the bump-like behavior of the
current and temperature dependence of the $\nu^{-}$ functions (see
Figs. 6 and 7 in Ref. 11 ), because $\nu^{-}_{i}$ and
$\nu^{-}_{a}$ functions, as it follows from {Eqs.}
(\ref{F32})-(\ref{F33}), give an important contribution to the odd
responses. The $\nu^{-}(j)$ and $\nu^{-}(\theta)$ curves for the
case of $\epsilon\ll1$ in fact are proportional to the derivatives
of the corresponding $\nu^{+}(j)$ and $\nu^{+}(\theta)$ curves,
which have a step-like behavior as a function of their arguments
(see Ref. 11 for the detailed discussion of this point and {Eq.}
(\ref{F17}) in this paper). As the $\rho^{-}_{\parallel}$ and
$\rho^{-}_{\perp}$ resistivities given by {Eqs.}
(\ref{F23})-(\ref{F24}) arise by virtue of the Hall effect, their
characteristic scale is proportional to $\epsilon\ll1$, as for
{Eqs.} (\ref{F19}) for purely anisotropic pins.

\begin{figure}
\includegraphics{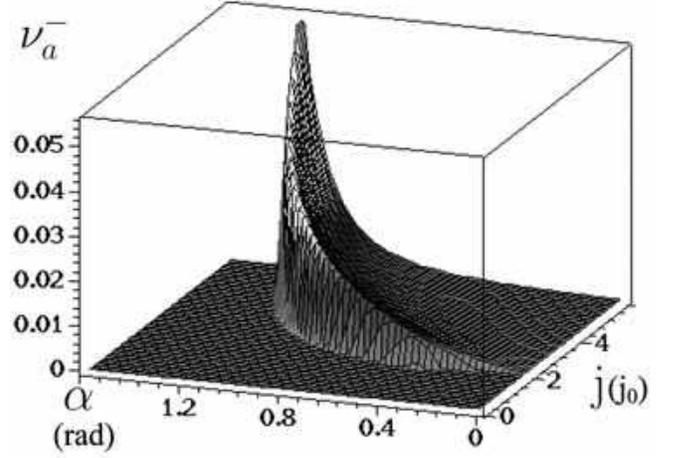}
\caption{The current-angle dependence of the function
$\nu_{a}^{-}(j,\alpha)$.} \label{fig11}
\end{figure}

\begin{figure}
\includegraphics{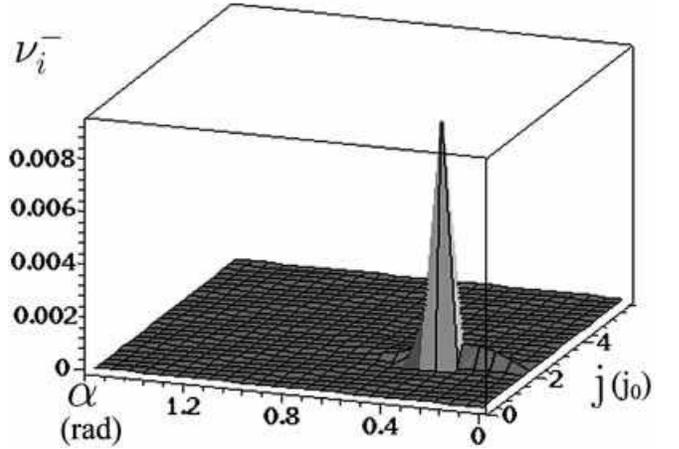}
\caption{The current-angle dependence of the function
$\nu_{i}^{-}(j,\alpha)$.} \label{fig12}
\end{figure}

The position of the characteristic peak in the $\nu^{-}_{i}$ and
$\nu^{-}_{a}$ functions is different for $q\neq1$, because
parameter $q$ determines the anisotropy of the critical current
densities for $i$- and $a$- pins. So, if $q$ is not very close to
the unity, the position of the $i$- and $a$- peaks cannot
coincide, and in this case the current and temperature odd
resistive dependences $\rho^{-}_{\parallel,\perp}$ can have a
bimodal behavior. For the $\rho^{-}_{\parallel}$ curves such
dependences will correspond to existence of the resistive "steps"
on the $\rho^{+}_{\parallel}$ curves (see Fig. 7), because for
$\epsilon\ll1$ we can consider the $\rho^{-}_{\parallel}$
dependences as derivatives of the $\rho^{+}_{\parallel}$ curves.
From this viewpoint it is easy to understand the previous
assertion in item E.2 of Sec. II that $\rho^{-}_{\parallel}$
includes  two terms (every proportional to the $\nu^{-}_{i}$ and
$\nu^{-}_{a}$, respectively) with similar signs.

Now we will discuss the $\nu_{a}^{-}$ and $\nu_{i}^{-}$ as a
function of ($j,\alpha$) and the parameter $q$ in detail. Really,
due to the smallness of the Hall constant, the $\nu_{a}^{-}$ and
$\nu_{i}^{-}$ tend to zero in the regions of the linear TAFF and
FF regimes of the $\nu_{a}$ and $\nu_{i}$ function, respectively.
The $\nu_{a}^{-}$ and $\nu_{i}^{-}$ functions have a sharp peak
(see Fig.~11,~12) in the region of sharp change of the $\nu_{a}$
and $\nu_{i}$ increasing (for $j\approx j_{ca}$ or $j\approx
j_{ci}$, respectively). With $\alpha$- and $q$-increasing the
width and the height of the $\nu_{a}^{-}$ maximum also increases
with simultaneous shift of the maximum to the higher current
densities due to the relation $j_{ca}(\alpha)\approx
q/\cos\alpha$. The $\nu_{i}^{-}$ peak is located in the angle
range $0<\alpha\lesssim\pi/4$, which corresponds to a change of
the angular dependence of the crossover current density
$j_{ci}(\alpha)$ from the angles $\alpha\gtrsim\pi/4$ to the
angles $\alpha\ll\pi/4$ (see. Fig.~12). The $\nu_{i}^{-}$ maximum
shifts to a smaller current densities with $q$-increasing due to
the $j_{ci}(\alpha)\approx1/\nu_{a}q\cos\alpha$,. The magnitudes
of the $\nu_{a}^{-}$ and $\nu_{i}^{-}$ are compete by an order of
magnitude for $0<\alpha\lesssim\pi/4$ and all $q$-values which
satisfy a condition $j_{ca}\approx j_{ci}$.

\begin{figure}
\includegraphics{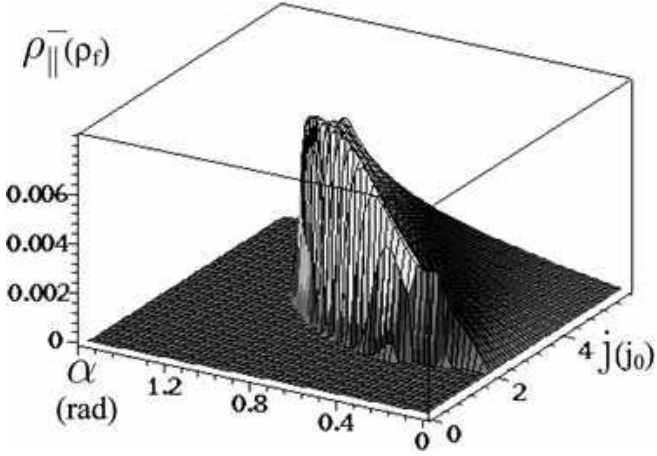}
\caption{The current-angle dependence of the odd longitudinal
magnetoresistivity $\rho_{\parallel}^{-}(j,\alpha)$.}
\label{fig13}
\end{figure}

\begin{figure}
\includegraphics{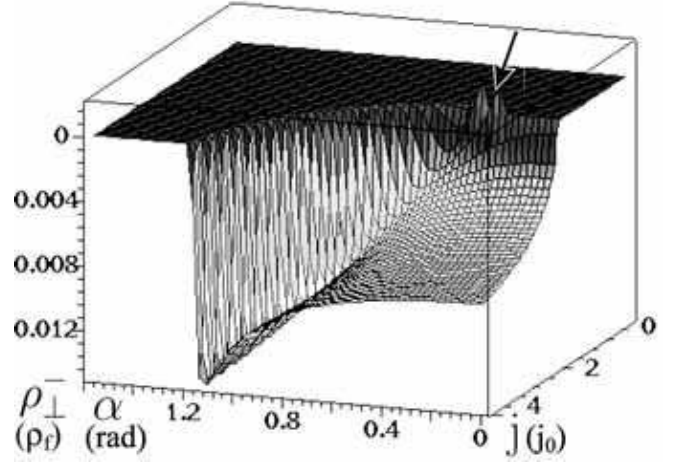}
\caption{The current-angle dependence of the odd transverse
magnetoresistivity $\rho_{\perp}^{-}(j,\alpha)$. The
characteristic minimum (which is shown by the arrow) is in the
region $0<\alpha<\pi/4$ and $j\approx j_{ci}(\alpha)$. The minimum
is shown as two neighboring minimums due to the step-like behavior
of the calculation. Pay attention to the inverted direction of the
axes in comparison with Fig.13.} \label{fig14}
\end{figure}

Now let us discuss a graphical presentation of the
{Eq.}~(\ref{F32}), which can be represented as
$\rho_{\parallel}^{-}=B_{1}+B_{2}$, where
$B_{1}=\nu_{i}^{-}\rho_{\parallel a}^{+}$, and
$B_{2}=\rho_{f}\nu_{a}^{-}\nu_{i}\cos^{2}\alpha$. Taking into
account that every factor in the $B_{1}$ and $B_{2}$ is positive
(see Figs. 5, 6, 11, 12), we can conclude that
$\rho_{\parallel}^{-}\geq0$  for all values of the $j,\alpha,q$.

Proceeding to the analysis of the $B_{1}$ and $B_{2}$
$(j,\alpha,q)$-behavior in details we consider first those
limiting cases in which $a$- or $i$- pinning is dominant i.e.
$\nu_{i}\approx1$ or $\nu_{a}\approx1$, respectively. If
$a$-pinning is dominant (i.e. for $q\gg1$), then
$\nu_{i}^{-}\rightarrow0$, and {Eq.}(\ref{F32}) has the form
$\rho_{\parallel}^{-}\approx\rho_{\parallel
a}^{-}=\nu_{a}^{-}\cos^{2}\alpha$. For the opposite case (i.e. for
$q<1$), conversely, $\nu_{a}^{-}\rightarrow0$, and
$\rho_{\parallel}^{-}\approx\nu_{i}^{-}\rho_{\parallel a}^{+}$.
The $\rho_{\parallel}^{-}(j,\alpha)$  graph presentation is
especially simple because it may be depicted with the aid of Figs.
5, 11, 12.

In the general case, i.e. for $q\approx1$, we should consider the
$B_{1}$ and $B_{2}$ separately because dominant type of pinning is
absent. The $B_{1}$ is proportional both $\nu_{i}^{-}$, which is
nonzero for $0<\alpha<\pi/4$ and $j(\alpha)\approx
j_{ci}(\alpha,q)$ (see Fig.~12), and the factor $\rho_{\parallel
a}^{+}$ with a graph, shown in Fig.~6. As a result, the $B_{1}$
has a sharp maximum for the $0<\alpha<\pi/4$ and $j(\alpha)\approx
j_{ci}(\alpha,q)$. The second term $B_{2}$ is proportional both
the factor $\nu_{a}^{-}$ and the factor $\nu_{i}\cos^2\alpha$. The
contribution of the first factor is maximal for
$\alpha\approx\pi/2$ and current densities $j(\alpha)\approx
j_{ca}(\alpha,q)$, whereas the $\nu_{i}\cos^2\alpha$ contribution
is maximal for $0<\alpha\lesssim\pi/4$ and $j(\alpha)\approx
j_{ci}(\alpha,q)$. Therefore these factors compete so that the
resulting maximum of $B_{2}$ shifts from $\alpha\approx\pi/2$ to
the $\pi/4\lesssim\alpha<\pi/2$. It is relevant to note that the
condition $j_{ci}(\alpha)<j_{ca}(\alpha)$ for $q\approx1$ and
$\pi/4\lesssim\alpha<\pi/2$ is always fulfilled. That is why the
maximal contribution of the $B_{2}$ is realized for
$j(\alpha)\approx j_{ca}(\alpha,q)$ because in this region of the
current densities the $\nu_{i}\rightarrow1$ for
$\pi/4\lesssim\alpha<\pi/2$. Therefore the $\rho_{\parallel}^{-}$
behavior is determined mainly by the $B_{2}$ behavior, and the
$B_{1}$ contribution is essential for $0<\alpha\lesssim\pi/4$ and
$j\approx j_{ci}(\alpha)$.

The $\rho_{\perp}^{-}(j,\alpha)$ dependence is the most
complicated. For the sake of simplicity the analysis we represent
the $\rho_{\perp}^{-}$ as a sum
$\rho_{\perp}^{-}=\rho_{f}[A_{1}+(A_{2}+A_{3})\sin2\alpha]$, where
$A_{1}=n\epsilon\nu_{a}\nu_{i}^{2}$, $A_{2}=\nu_{a}^{-}\nu_{i}/2$,
$A_{3}=-\nu^{-}_{i}(1-\nu_{a})/2$. First we consider the limiting
cases of purely isotropic or anisotropic pinning
($\nu_{a}\rightarrow1$ or $\nu_{i}\rightarrow1$, respectively).
For $i$-pinning we have
$\rho_{\perp}^{-}=\rho_{f}n\epsilon\nu_{i}^{2}$, from which
follows (Vinokur et al.$^{17}$) a scaling relation
$\rho_{\perp}\sim\rho_{\parallel}^2$. For the case of purely
anisotropic pinning
$\rho_{\perp}^{-}=\rho_{f}\{n\epsilon\nu_{a}+(\nu_{a}^{-}\sin2\alpha)/2\}$,
and the scaling relation is $\rho_{\perp}\sim\rho_{\parallel}$
(see also Ref. 16).

Now we consider every term in the $\rho_{\perp}^{-}(j,\alpha)$ in
detail. The $A_{1}$ contribution can be reduced in fact to the
multiplication of the graph in Fig. 3 by the graph in Fig. 5
squared; the result is essentially nonzero for $j\gtrsim
j_{ca}(\alpha,q)$. The $A_{2}$ contribution was described above
(see the $B_{2}$ term in the $\rho_{\parallel}^{-}$ without taking
into account the $\cos^2\alpha$ anisotropy). Note also that both
terms ($A_{1}$ and $A_{2}$) are positive for $n\epsilon>0$. The
$A_{3}$ behavior is of great interest because the $A_{3}<0$ for
$n\epsilon>0$. Let us consider the cases $q>1$ and $q<1$, which
correspond to the $a$-, or $i$-pinning domination, respectively.
Then, for $\alpha<\pi/4$:

a) for $q<1$ we have $j_{ci}(\alpha)>j_{ca}(\alpha)$ and the sharp
maximum of the $\nu^{-}_{i}$ is suppressed by the factor
$(1-\nu_{a})\rightarrow0$. As a result, the $A_{3}$ contribution
can be ignored.

b) for $q\geq1$ the opposite inequality follows, i.e.
$j_{ci}(\alpha)<j_{ca}(\alpha)$. Then for $j\approx
j_{ci}(\alpha)$ the $A_{3}$ term is dominant because $\nu_{a}\ll1$
and $\nu_{i}^{-}\rightarrow n\epsilon$ in this $(j,\alpha)$-region
(see Fig.~3 and Fig.~5). As a result, the
$\rho_{\perp}^{-}(j,\alpha,q)$ change the sign for $j\approx
j_{ci}(\alpha)$ and $0<\alpha\lesssim\pi/4$. Since the scale of
the $\nu_{i}^{-}\ll\nu_{i}$, the amplitude of the minimum is small
in comparison with the $\rho_{\perp}^{-}$ magnitude.

Thus, a competition of the $a$- and $i$-pinning leads to the
qualitatively important conclusion that the $\rho_{\perp}^{-}$ can
change its sign at a certain range of $(\alpha,j,q)$-values,
namely for $j\approx j_{ci}(\alpha,j,q)$, $0<\alpha\lesssim\pi/4$,
and $q>1$.

\subsection{Resistive response in a rotating current scheme.}

\subsubsection{Polar diagram.}

An experimental study of the vortex dynamics in $\rm {
YBa_{2}Cu_{3}O_{7-\delta}}$ crystals with unidirectional twin
planes was recently done using a modified rotating current
scheme$^{4,5}$. In that scheme it was possible to pass current in
an arbitrary direction in the $ab$ plane of the sample by means of
four pairs of contacts placed in the plane of the sample. Two
pairs of contacts were placed as in the conventional four-contact
scheme, and the other two pairs were rotated by $90^{\circ}$ with
respect to the first (see the illustration in Fig. $1$ of Ref. 4).
By using two current sources connected to outer pair of contacts,
one can continuously vary the direction of the current transport
in the sample. By simultaneously measuring the voltage in the two
directions, one can determine directly the direction and magnitude
of the average velocity vector of the vortices in the sample as a
function of the direction and magnitude of the transport current
density vector. This made it possible to obtain the angular
dependence of the resistive response on the direction of the
current with respect to the pinning planes on the same sample. The
experimental data$^{4,5}$ attest to the anisotropy of the vortex
dynamics in a certain temperature interval which depends on the
value of the magnetic field. A rotating current scheme was
used$^4$ to measure the polar diagrams of the total
magnetoresistivity $\rho(\alpha)$, where
$\rho=(\rho^{2}_{x}+\rho^{2}_{y})^{1/2}$ is the absolute value of
the magnetoresistivity, $\rho_{x}$ and $\rho_{y}$ are the $x$ and
$y$ components of the magnetoresistivity in an $xy$ coordinate
system, and $\alpha$ is the angle between the current direction
and the $oy$ axis (parallel to the channels of the $a$-pinning
centers). In the case of a linear anisotropic response the polar
diagram of the resistivity is an ellipse, as can easily be
explained. In the case of a nonlinear resistive response the polar
diagram of the resistivity is no longer an ellipse and has no
simple interpretation.

In this subsection we carry out a theoretical analysis of the
polar diagrams of the magnetoresistivity $\rho$ in the general
nonlinear case in the framework of a stochastic model of $a+i$
pinning. This type of angular dependence $\rho(\alpha)$ is
informative and convenient for theoretical analysis. For a sample
with specific internal characteristics of the pinning (such as
$q$, $\varepsilon_{a}$, $\varepsilon_{i}$, and $\kappa$) at a
given temperature and current density the function $\rho(\alpha)$
is contained by the resistive response of the system in entire
region of angles $\alpha$ and makes it possible to compare the
resistive response for any direction of the current with respect
to the direction of the planar pinning centers. In addition, in
view of the symmetric character of the $\rho(\alpha)$ curves,
their measurements makes it possible to establish the spatial
orientation of the system of the planar pinning centers with
respect to the boundaries  of the sample if this information is
not known beforehand.

\begin{figure}
\includegraphics{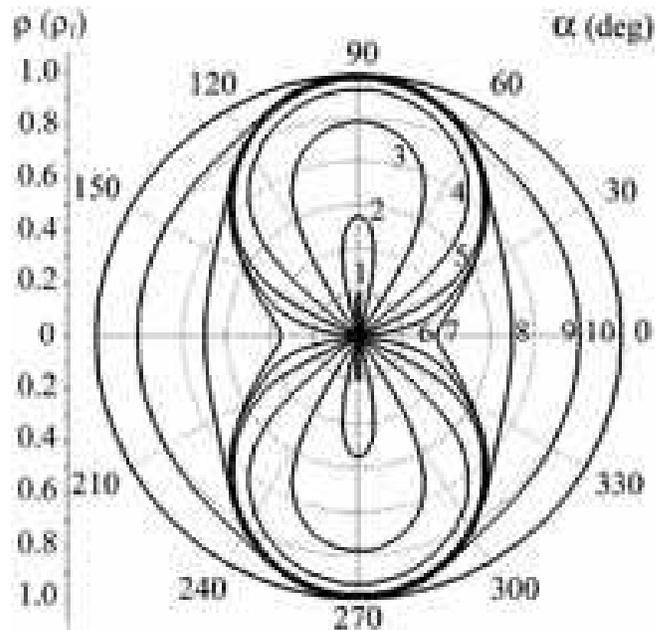}
\caption{Series of graphs of the function $\rho(\alpha)$ for a
sequence of the parameter $j$: 0.63 (1), 0.65 (2), 0.75 (3), 1.00
(4), 1.50 (5), 1.92 (6), 2.00 (7), 2.50 (8), 4.00 (9), 20.0 (10)
for $\varepsilon_{a}=1$.} \label{fig15}
\end{figure}

\begin{figure}
\includegraphics{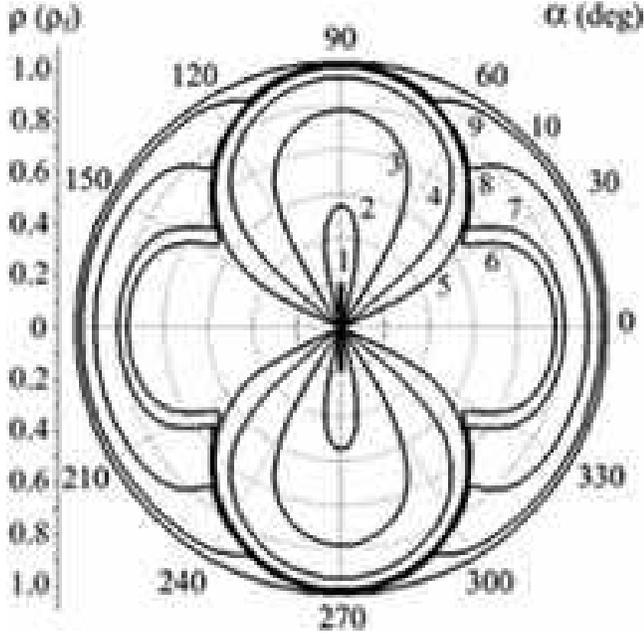}
\caption{Series of graphs of the function $\rho(\alpha)$ for a
sequence of the parameter $j$: 0.63 (1), 0.65 (2), 0.75 (3), 1.00
(4), 1.50 (5), 1.92 (6), 2.00 (7), 2.50 (8), 4.00 (9), 20.0 (10)
for $\varepsilon_{a}=0.1$.} \label{fig16}
\end{figure}

Now for analysis of the $\rho(\alpha)$ curves we imagine that
vector  \textbf{j} is rotated continuously from an angle
$\alpha=\pi/2$ to $\alpha=0$. The characteristic form of the
$\rho(\alpha)$ curves will obviously be determined by the sequence
of dynamical regimes through which the vortex system passes as the
current density vector is rotated. By virtue of the symmetry of
the problem, the $\rho(\alpha)$ curves can be obtained in all
regions of angles $\alpha$ from the parts in the first quadrant.

We recall that in respect to the two systems of pinning centers it
is possible to have the linear TAFF and FF regimes of vortex
dynamics and regimes of nonlinear transition between them. The
regions of nonlinear transitions are determined by the
corresponding values of the crossover current densities
$j_{ci}(\alpha,q)$ and $j_{ca}(\alpha,q)$.

Now let us consider the typical $\rho(\alpha)$ dependences which
are presented in Fig. 15 and 16 for a sequence of a current
density magnitude. We remind that the polar diagram graphs
represented below are constructed, as the previous graphs in Figs.
3-7, 9-14 for the next values of parameters: $q=1.6$ (i.e. for the
case with dominant $a$-pins), $\kappa=0.25$, $\theta=0.003$,
$\epsilon=0.01$, $\varepsilon_{i}=0.1$, $\varepsilon_{a}=1$
(Fig.~15), and $\varepsilon_{a}=0.1$ (Fig.~16). Note that
$\rho(\alpha)$ is the product of two multipliers: one is the
$\nu_{i}(f_{i})$ dependence, which was earlier studied in Fig. 4
of item C.1 of Sec. III, and other is the
$\sqrt{\sin^2\alpha+\nu_{a}^2\cos^2\alpha}$ factor, which
qualitative behavior is close to the $\rho_{\parallel
a}^{+}(j,\alpha)$ dependence (see Fig. 6 in item C.2 of Sec. III).

Let us analyze the $\rho(\alpha)$ behavior for the series of
values of the current density $j$. When the angle $\alpha$ changes
from $0$ to $\pi/2$ the function $\rho(\alpha)$ grows
monotonically from $\rho(0)=\nu_{a}(j/q)\nu_{i}(jq\nu_{a}(j/q))$
to $\rho(\pi/2)=\nu_{i}(jq)$. In Fig.~15 curves 1-6 of the
function $\rho(\alpha)$ have the shape of the 8-figure drawn along
$ox$-axis (strongly elongated for the curves 1, 2).

This anisotropy can be determined by the relation of the
magnitudes of the half-axis at the direction $\alpha=\pi/2$ to the
transverse half-axis for any fixed magnitude of the current
density. The curves 1-6 of the $\rho=\rho(\alpha)$ graph has the
8-form elongated along $ox$-axis. It is caused by the step-like
behavior of the $\nu_{i}$-function, corresponding for the curves
1, 2 to the crossover from the TAFF to the FF regime. That is why
the magnitude of the $\rho(\pi/2)$ for the curve 2 is rather
greater than for the first one. With $\alpha$-increasing the
$\nu_{i}$-function is in the TAFF-region (see Fig.5), which
provocates the $\rho(\alpha)\ll1$ in the case where the condition
$j<j_{ci}(\alpha)$ is satisfied. Therefore, with $j$-increasing
the magnitude of the angle $\alpha$, which separates the TAFF and
the FF regions of the $\nu_{i}$-function at a fixed value of the
current density, decreases to the $\alpha\ll1$.

As the $\nu_{i}(j,\alpha=\pi/2)$ is in the FF region (i.e.
$j\gtrsim~j_{ci}(\alpha=\pi/2)$), so the anisotropy of the 8-curve
decreases for curves 3-6. The $\rho=\rho(\alpha)$ behavior of the
curves 5-6 is more isotropic in the region $\alpha\ll\pi/4$ than
behavior of the curves 1-4. If the condition $j>j_{ca}(0)$ is
satisfied, an appearance of the nonzero resistance in
corresponding region follows. Its magnitude is smaller than
$\rho(\pi/2)$ for the curves 7, 8, 9 and practically is equal to
the $\rho(\pi/2)$ for the curve 10. Note, that for the
$\alpha=0,\pi$ and $j_{ca}<j\lesssim3j_{ca}/2$ one can see the
minimum, which decreases with $j$-increasing and disappears in the
case where the condition $j\gtrsim3j_{ca}/2$ is satisfied. So, for
large magnitudes of the current densities the $\rho(\alpha)$
behavior becomes more isotropic.

It is necessary to pay attention for the $\rho=\rho(\alpha)$
behavior in the case where $\varepsilon_{a}=0,1$ (see Fig. 16) for
the same series of the magnitudes of the current densities. The
behavior of the curves 6, 7, 8, 9 differs from the above-mentioned
case, but the behavior of the curves 1-5, 10 retains the same.
This fact is caused by the influence of the parameter
$\varepsilon_{a}$ on the $\nu_{i}$ behavior only in the area of
its sharp step-like behavior at the $j\simeq j_{ca}(\alpha)$.
Note, that the $\nu_{i}$ contribution is dominant in the region
$\alpha\ll1$ as well as the above-mentioned anisotropy of the
$\rho_{\perp a}^{+}(j,\alpha)$ (see Fig. 5 and Fig. 7). As
decreasing of the $\varepsilon_{a}$ causes the more narrow
crossover from the TAFF- to the FF-regime, the $\nu_{i}(\alpha)$
has a minimum at fixed magnitude of the current density. The
magnitude of this minimum decreases with the $j$-increasing and
the minimum shifts from the $\alpha\simeq\pi/4$ to the
$\alpha\simeq\pi/2$. The influence of the parameter $q$ acts on
the crossover current densities $j_{ci}$ and $j_{ca}$ only
quantitatively, but does not change an evolution of the curves
1-10 qualitatively.

\subsubsection{$\Theta_{E}(\alpha)$-dependence.}

Let us examine theoretically in our model a new type of the
experimental dependence, recently studied$^4$ for
$\Theta_{E}(\alpha)$, where $\Theta_{E}$ is the angle between
$\mathbf{j}$-vector and the electric field vector $\mathbf{E}$
measured at fixed values of the current density and temperature.
Taking into account that in the \emph{xy} coordinate system the
magnetoresistivity components are
$\rho_{x}=\rho_{xx}\sin\alpha=\nu_{i}(F_i)\sin\alpha$,
$\rho_{y}=\rho_{yy}\cos\alpha=\nu_{i}(F_i)\nu_{a}(F_a)\cos\alpha$,
we obtain the following simple relation:
$\tan\Theta_{E}(\alpha)=\rho_{x}/\rho_{y}=\tan\alpha/\nu_a(F_a)$,
or
\begin{equation}
\label{F36}\Theta_{E}(\alpha)=\arctan(\tan\alpha/\nu_a(F_a)).
\end{equation}
Note, that the $\nu_{i}$ term, describing the $i$-pinning, is
absent in {Eq.}(\ref{F36}). Then it follows from the latter  that
the
$\nu_{a}(j,\alpha,\theta)=\tan\alpha/\tan\Theta_{E}(j,\alpha,\theta)$,
i.e. the $\nu_{a}(j,\alpha,\theta)$ function can be found from the
experimental dependence $\Theta_{E}(\alpha)$. Unfortunately, the
dependence $\Theta_{E}(\alpha)$ for the series of the temperature
values was experimentally found$^{4}$ so far only for the
FF-regime (see. Fig.~2 in Ref.~4). The $\Theta_{E}(j,\alpha)$
dependence is presented in Fig.~17. It shows all changes in the
$\Theta_{E}(j,\alpha)$ behavior also for the TAFF-regime.

\begin{figure}
\includegraphics{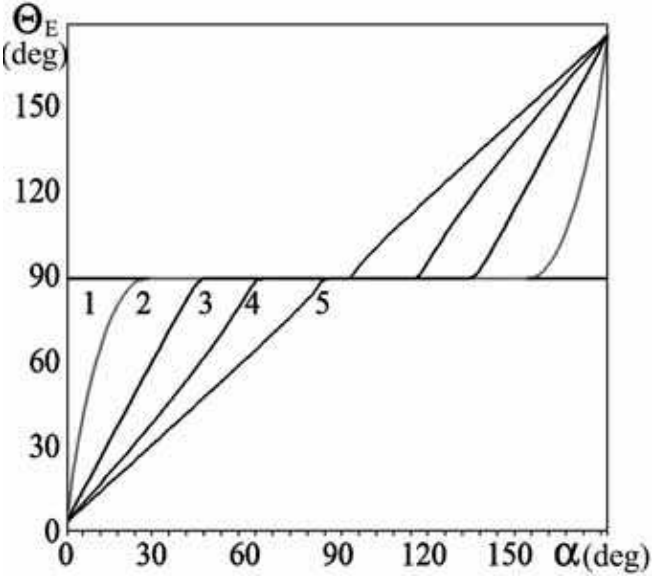}
\caption{Series of graphs of the function $\Theta_{E}(\alpha)$ for
a sequence of the parameter $j$: 1 (1), 1.7 (2), 2.2 (3), 3.5 (4),
20 (5) for $T=8K$.} \label{fig17}
\end{figure}

Let us analyze the {Eq.}~(\ref{F36}) in detail. The
$\Theta_{E}(j,\alpha)$ is the odd function of the angle $\alpha$,
and its magnitude increases monotonically with the
$\alpha$-increasing for all values of the $j$ due to the
monotonical decreasing of the $\nu_{a}(j,\alpha)$ function (see.
Fig. 17). It follows from {Eq.}~(\ref{F36}) that the period of the
function $\Theta_{E}(\alpha)$ is equal to $\pi$. One more
important limiting case is realized for $\nu_{a}\approx1$,which
corresponds to the limit of isotropic pinning. Depending on the
inequality between the $j$ magnitude and the crossover current
density $j_{ca}(\alpha)\approx q/\cos\alpha$, one can separate two
regions where the $\Theta_{E}(j,\alpha)$ behavior is qualitatively
different. If $\emph{A}$ is the argument of the arctangent
function in {Eq.}~(\ref{F36}), then in that region $j,\alpha,q$,
where the inequality $j\gtrsim~j_{ca}(\alpha)$ is true (the FF
regime for $\nu_{a}(j,\alpha)$, see also Fig.~3), the magnitude of
the $\Theta_{E}\approx\emph{A}$ as $\emph{A}\ll1$. And for the
case $j\lesssim~j_{ca}$ (the TAFF regime of the
$\nu_{a}(j,\alpha)$) the value
$\Theta_{E}\approx\pi/2-\emph{A}^{-1}$, as $\emph{A}\gg1$.

Note, that the parameter $\varepsilon_{a}$ influences the
$\Theta_{E}(j,\alpha)$ by changing the character of the step-like
crossover of the $\nu_{a}(j,\alpha)$ (the smaller the
$\varepsilon_{a}$, the sharper the crossover). The value of the
parameter $q$, as well as above-mentioned, determines the
magnitude of the $j_{ca}(\alpha)$  (and, therefore the position of
the boundaries in $j$ of the regions of quite different
$\Theta_{E}(\alpha)$ behavior) at fixed $\alpha$.

\subsubsection{Critical current density anisotropy.}

Under the critical current density we mean the current density,
which corresponds to the electric field strength on the sample
$E=1 \mu V/cm$. Let us determine the $j_{c}(\alpha)$ behavior
graphically by crossing the
$E_{\parallel}^{+}=j\rho_{\parallel}^{+}(j)$ graph and the plain
$E=E_{c}$ in the polar coordinates. For all angles $\alpha$ the
point of crossing for these graphs determines the critical current
density magnitude for the defined direction, and the crossing line
of the graphs presents the dependence $j_{c}(\alpha)$.

Let us remind the reader that as in above-mentioned sections, in
the nonlinear law $E_{\parallel}^{+}=j\rho_{\parallel}^{+}(j)$ we
measure $j$ and $\rho$ in the values of the
$j_{0}=cU_{0}/\Phi_{0}dh$ and $\rho_{f}=\rho_{n}B/B_{c2}$,
respectively. That is why the $E$ magnitude we have to measure in
the $E_{0}=j_{0}\rho_{f}$. As well as in item B of Sec. III we use
the data from Ref.~8, where for the niobium samples
$\rho_{n}\approx5,5\cdot10^{-6}$ Ohm$\cdot$cm, $B\approx150$ Gs,
$B_{c2}\approx$ 17 kGs,
$\rho_{f}\approx5\cdot10^{-8}$Ohm$\cdot$cm, $U_{0}=2500 K$, and
$d=2.5\cdot10^{-6} cm$.

Therefore, $E_{0}\approx6\cdot10^{-4}V/cm$, and for $E_{c}=1 \mu
V/cm$ we have to cross the dimensionless
$\rho_{\parallel}^{+}(j)\cdot j$ graph by the plain
$E\approx0.002$.

\begin{figure}
\includegraphics{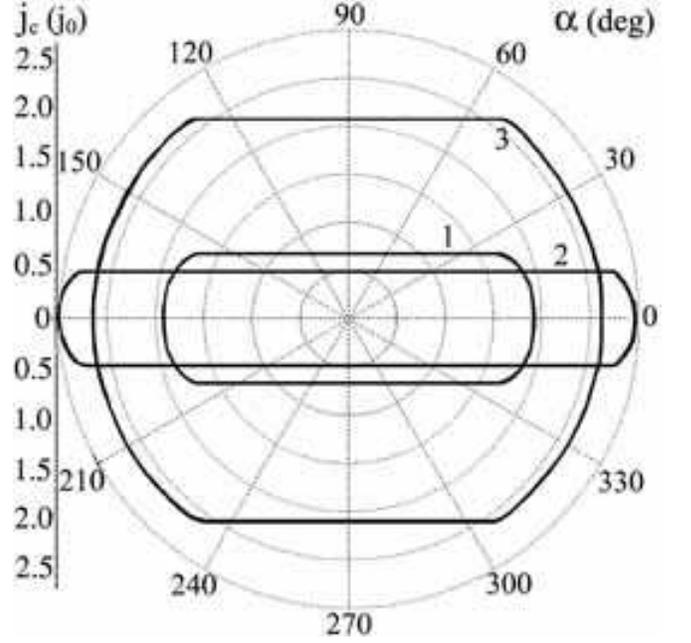}
\caption{Series of graphs of the function $j_{c}(\alpha)$ for the
parameter pairs: $E_c=0.002$, $q$=1.6 (1); $E_c=0.002$, $q$=3 (2);
$E_c=2$, $q$=1.6 (3).} \label{fig18}
\end{figure}

Now we will discuss the $j_{c}(\alpha)$ as a function of
$\alpha,q,E_c$, and $\varepsilon$ in detail. The
$j_{c}(\alpha)$-anisotropy can be determined by the relation of
the magnitudes of the half-axis at the direction $\alpha=0$ to the
transverse half-axis for any fixed magnitude of the parameters
$q$, $E_c$. The $j_{c}(\alpha)$ decreases monotonically from
$j_{c}(0)$ with $\alpha$-increasing and has a minimum for
$\alpha=\pi/2$. It is caused by the fact that, as it was shown in
item C.1 of Sec. II, the $a$-pinning (with high values of the
$j_{ca}$ for $q>1$) does not influence the $i$-pinning for
$\alpha=\pi/2$. Therefore, the inequality for the crossover
current densities $j_{ci}(\alpha)<j_{ca}(\alpha)$ for $q>1$ leads
to the corresponding inequality for the critical current densities
$j_{c}(0)<j_{c}(\pi/2)$.

The $q$ influences the $j_{c}(\alpha)$ behavior (as in item D.1 of
Sec. II) only quantitatively: with $q$-increasing the ratio
$j_{c}(0)/j_{c}(\pi/2)$ grows and visa versa. It is caused by the
$j_{c}(0)$-increasing and $j_{c}(\pi/2)$-decreasing due to the
$\alpha$-behavior of the corresponding crossover current densities
$j_{ci}(\alpha)$ and $j_{ca}(\alpha)$. The smaller the
$\varepsilon_a$, the sharper the crossover between the
$j_{c}(\alpha)$ regions of slowly and quickly decreasing as a
function of the $\alpha$. With $E_c$-increasing the nonlinear law
$E_{\parallel}^{+}=\rho_{\parallel}^{+}(j)j$ is satisfied for the
larger values of the current density.

That is why with $\alpha$-increasing from $0$ to $\alpha^{\ast}$
values (for which the condition
$\tan^{2}\alpha^{\ast}\ll\nu_{a}^{2}(j,\alpha^{\ast})$ is
satisfied) the $\nu_{a}$-function is in the FF regime and
$j_{c}(\alpha)$ decreases slowly. When the condition
$\alpha>\alpha^{\ast}$ is true the $\nu_{a}$-function has a
step-like crossover from the FF to the TAFF regime and
$j_{c}(\alpha)$ decreases quickly.

So, the $\alpha^{\ast}$ behavior as a function of the parameters
$q$ and $E_c$ is qualitatively different: it increases with
$E_c$-increasing and decreases with $q$-increasing. On the
increase of the $E_c$ by the several orders of magnitude the
$j_{c}(\alpha)$ curve degenerates into a circumference due to the
isotropization of the $j_{ci}(\alpha)$ and $j_{ca}(\alpha)$
behavior for the high $j$-values. Otherwise, with $E_c$-decreasing
the $j_{c}(\alpha)$ curve degenerates into a narrow loop, because
the $j_{ci}(\alpha)$ and $j_{ca}(\alpha)$ behavior for a small $j$
is very anisotropic.

\section{CONCLUSION.}

In the present work we have theoretically examined the strongly
nonlinear anisotropic two-dimensional single-vortex dynamics of a
superconductor with coexistence of the anisotropic washboard PPP
and isotropic pinning potential as function of the transport
current density $j$ and the angle $\alpha$ between the direction
of the current and PPP planes at a fixed temperature $\theta$.

The experimental realization of the model studied here can be
based on both naturally occurring$^{2-5}$ and artificially
created$^{6-8}$ systems with $i+a$ pinning structures. The
proposed model has made it possible for the first time (as far as
we know) to give a consistent description of the nonlinear
anisotropic current- and temperature-induced depinning of vortices
for an arbitrary direction relative to the anisotropy of the
washboard PPP. In the framework of this model one can successfully
analyze theoretically certain observed resistive responses which
are used for studying anisotropic pinning in a number of new
experimental techniques$^{4}$ (the polar diagram of
$\rho(\alpha)$, the $\Theta_{E}(\alpha)$ curve described by
formula {Eq.}~(\ref{F36})) as well as new Hall responses specific
for the $i+a$ pinning problem.

A quantitative description of the anisotropic nonlinear resistive
properties of the problem under study is done in the framework of
the stochastic model on the basis of the Fokker-Planck approach.
The main nonlinear components of the problem are the anisotropic
$\nu_{a}(F_{a})$ and isotropic  $\nu_{i}(F_{i})$ probability
functions for the vortices to overcome the potential barriers of
$a$- and $i$-pinning centers under the action of anisotropic
motive forces $F_{a}$ and $F_{i}$, respectively. The latter
include both the "external" parameters $j,\alpha,\theta$ and the
"internal" parameters $q,\varepsilon_i,\varepsilon_a$ which
describe the intensity and anisotropy of the pinning. As can be
seen from {Eqs.} (\ref{F30})-(\ref{F33}), the magnetoresistivities
$\rho_{\parallel,\perp}^\pm(j,\alpha,\theta)$ are, in general,
nonlinear combinations of the experimentally measured $\nu_{i}$
and $\nu_{a}$ functions ($\nu_{i}$ can be measured independently
from the $\rho_{\parallel,\perp}^+(\alpha=\pi/2)$, see {Eq.}
(\ref{F30}) and $\nu_{a}$ - from the $\Theta_{E}(\alpha)$, see
{Eq.} (\ref{F36})).

Therefore, the nonlinear (in $j$) resistive behavior of the vortex
system can be caused by factors of both an anisotropic and
isotropic pinning origin. It is important to underline that
whereas the structure of the $\nu_{a}(F_{a})$ and $F_{a}$ is the
same as for purely $a$-pinning problem, the structure of the
$\nu_{i}(F_{i})$ and $F_{i}$ is strongly different from the
structure of the purely $i$-pinning problem due to the fact that
$F_{i}$, as motive force of the ($i+a$)-problem, is nonlinear and
anisotropic (see {Eqs.}~(\ref{F34})-(\ref{F35})) and
Figs.~3,~4,~5).

Two main new features appear due to the introduction of the
isotropic $i$-pins into the initially anisotropic $a$-pinning
problem. First, unlike the stochastic model of uniaxial
anisotropic pinning studied previously$^{10,11}$, where the
critical current density $j_c$ is indeed equal to zero for all
directions (excepting $\alpha=0$) due to the guiding of vortices,
in the given $i+a$ model the anisotropic critical current density
$j_c(\alpha)$ exists for all directions because $i$-pins "quench"
the guiding of vortices in the limit $(j,\theta)\rightarrow0$.
Second, the Hall resistivity response functions
$\rho_{\perp}^{-}(j,\alpha)$ can have a change of sign in a
certain range of $(j,\alpha,q)$ (at fixed dimensionless Hall
constant $\epsilon=\alpha_H/\eta$ and the dimensional Hall
conductivity $\delta=n\epsilon/\rho_f)$, whereas the sign of the
$\rho_{\parallel}^{-}(j,\alpha)$ does not change.

It should be noted that recently$^{8}$ the nonlinear (in~$\theta$)
anisotropic longitudinal and transverse resistances of Nb films
deposited on facetted sapphire substrates were measured at
different angles $\alpha$ between $\mathbf{j}$ and facet ridges in
a broad range of temperature and relatively small magnetic field
$\mathbf{H}$. The experimental data were in good agreement with
the theoretical model described here. The measured
$\rho_\parallel^+(\theta,\alpha)$ dependences can be fitted using
the probability functions $\nu_a$ and $\nu_i$ in the form proposed
here (see {Eq.}~(\ref{F29})) with the anisotropic and isotropic
pinning potential given by {Eq.}~(\ref{F28}). The periods and
depths of the potential wells were estimated from the experimental
data$^8$ and were used here (see Sec.III) for the theoretical
analysis of different types of nonlinear anisotropic
$(j,\alpha)$-dependent magnetoresistivity responses, given by
{Eqs.}~(\ref{F30})-(\ref{F33}), in the form of graphs (see
Figs.~3-18). Whether these theoretical results can explain a new
portion of the $(j,\alpha)$-dependent $i+a$ resistivity data
measured at $\theta=const$ (in particular, for the samples
investigated earlier$^8$ at small current densities) remains to be
seen.

\end{document}